\documentclass[aps,pra,twocolumn,nopacs,floatfix]{revtex4-2}
\usepackage{epsfig}
\usepackage{graphicx}
\usepackage{dcolumn}
\usepackage{amsthm,amsmath}
\usepackage{comment}
\usepackage[colorlinks=true, allcolors=blue]{hyperref}
\usepackage{xcolor}
\usepackage{mathtools}
\usepackage{orcidlink}
\usepackage{amsmath}
\usepackage[normalem]{ulem}
\usepackage{bm}
\usepackage{soul}

\begin{document}

\title{Comparative Analysis of Mg$^+$ Properties using Multiconfiguration Dirac-Hartree-Fock and Relativistic Coupled-cluster Methods}

\author{B. K. Sahoo~\orcidlink{0000-0003-4397-7965}}
\email{bijaya@prl.res.in}
\affiliation{
Atomic, Molecular and Optical Physics Division, Physical Research Laboratory, Navrangpura, Ahmedabad 380009, India}  

\author{Per J\"onsson~\orcidlink{0000-0001-6818-9637}}
\email{per.jonsson@mau.se}
\affiliation{
Malmö University, Faculty of Technology and Society (TS), Department of Materials Science and Applied Mathematics (MTM)}

\author{Gediminas Gaigalas~\orcidlink{0000-0003-0039-1163}}
\affiliation{
Institute of Theoretical Physics and Astronomy, Faculty of Physics, Vilnius University, Saul\.{e}tekio Ave. 3, LT-10257 Vilnius, Lithuania}

\date{\today}

\begin{abstract}
We demonstrate behaviors of correlation effects in the calculations of atomic properties through two commonly employed many-body methods; namely multiconfiguration Dirac-Hartree-Fock (MCDHF) and relativistic coupled-cluster (RCC) methods. Particularly, we have bench-marked excitation energies, electric dipole (E1) matrix elements, magnetic dipole hyperfine structure constants ($A_{hf}$), and isotope shift (IS) constants in the singly ionized magnesium (Mg$^+$) systematically at different levels of approximation of both methods. We have also estimated the E1 polarizability of the ground state and lifetimes of the excited states using the E1 matrix elements from both methods. All these results are compared with the experimental values wherever available. We find that the computed results agree well with each other with a few exceptions; particularly the $A_{hf}$ and IS constants from the RCC method are found to agree with the measurements better. This comparison analysis would be useful in evaluating the above-discussed properties in other atomic systems using the MCDHF and RCC methods more reliably.
\end{abstract}

\maketitle

\section{Introduction}

The usefulness of high-precision calculations of atomic properties is many-fold. Some of the typical applications of atomic calculations are a comparison between the calculated and experimental results that can demonstrate the potential of a many-body method to produce accurate results, high-precision calculations can help guiding experiments to carry out measurements in the right direction and they can be used to predict various systematic effects during measurements, etc. to name a few. However, atomic calculations for situations like to probe atomic parity violations (APVs) \cite{bijaya-pnc}, electric dipole moments (EDMs) due to simultaneous violations of parity and time reversal symmetries \cite{bijaya-xe, bijaya-cs, bijaya-fr}, possible temporal and spatial variations of the fine-structure constant ($\alpha_e$) \cite{bijaya-dillip}, local Lorentz symmetry invariance (LLI) \cite{bijaya-ca}, extraction of nuclear charge radii and signature of a vector boson particle from the isotope shift (IS) studies \cite{bijaya-review}, and inferring nuclear moments from the measured hyperfine structure constants are equally important with the high-precision atomic experiments \cite{PhysRevLett.87.133003,bijaya-sr, bijaya-in}. This demands a better understanding of atomic many-body methods and their capabilities to produce accurate results. 

At present, the relativistic coupled-cluster (RCC) theory and the multiconfiguration Dirac-Hartree-Fock and relativistic configuration interaction (MCDHF/RCI) methods are well-known many-body methods that are commonly employed to estimate atomic properties accurately \cite{bijaya-review, das, grant2007relativistic, atoms11010007}. It has been noticed earlier that calculations of some of the properties from both the methods agree, and some of them differ substantially \cite{bijaya-pradeep, Lifetime, polz, Leonid}. Many times, the differences in the results are attributed to the use of finite-size basis functions and approximations made in the methods. As long as experimental results are available, it may be possible for someone to explain the reasons for such differences. However, it would be necessary to decide the reliabilities of calculations from both the methods for quantities that are required in the aforementioned analyses, like enhancement factors due to atomic APV, EDM, variation of the fine-structure constant and LLI as well as IS constants. For these quantities, one cannot get experimental results, so their accurate determinations depend entirely on the choice of a many-body method.

In order to demonstrate how calculations of properties with different radial dependencies converge at different approximations in the MCDHF/RCI and RCC methods, we analyze excitation energies, magnetic dipole hyperfine structure constants ($A_{hf}$), electric dipole (E1) amplitudes and IS constants of several low-lying states of $^{25}$Mg$^+$ in this work by employing both the methods. In fact, we had demonstrated in an earlier work the behavior of electron correlation effects in the determination of properties in Ca$^+$ described by operators with different radial and angular momentum dependencies considering lower-order many-body methods and RCC method \cite{bijaya-pradeep}. In the RCC method, calculations are started with a single reference Dirac-Hartree-Fock (DHF) wave function and the calculations are carried out by considering the singles and doubles approximated RCC (RCCSD) method and the singles, doubles, and triples approximated RCC (RCCSDT) method.     

The non-relativistic approximation of the RCC (CC) theory has been applied to atoms \cite{bartlett,deyonker,nataraj,eliav}, molecules \cite{cizek,bartlett,crawford,prasana}, condensed matter systems \cite{bishop01,bishop} and nuclei \cite{kowalski,hagen}. It is currently one of the leading quantum many-body methods and has been referred to as the gold standard for treating electron correlation \cite{bartlett, crawford, bishopbook1, bishopbook}. It is straightforward to apply the RCC methods for the evaluation of energies but not for other properties. Thus, a number of variant approaches in the RCC theory framework are developed for determining atomic properties in which both the ket and bra states are evaluated separately \cite{bartlett,bishop01,bishop,bijaya-ncc,bijaya-ar}. We consider here the bra state as the complex conjugate (c.c.) of the RCC theory to evaluate the E1 matrix elements and $A_{hf}$ values. However, we determine the IS constants through the finite-field (FF) and analytical response (AR) approaches in the RCC theory framework \cite{bijaya-review, bijaya-ar}. The FF approach includes orbital relaxation effects, while the AR approach avoids numerical differentiation of the total energies with respect to the perturbative parameter in the evaluation of the IS constants \cite{bijaya-k}.

The MCDHF/RCI method is known to be very versatile and easier to apply to any open-shell systems \cite{atoms5020016}. Results from full RCI can be used to test the RCC calculations, but considering full RCI beyond a boron-like system is almost impossible. On the other hand, a truncated RCI would develop a size-extensitivity problem for heavier atomic systems \cite{bartlett, Lindgren1986}. In a lighter system like Mg$^+$, the RCI method is expected to produce accurate results. Unlike the RCC method, all possible excitation configurations in the RCI method appear in the linear form. Therefore, it is absolutely required to consider large number of configuration state functions (NCSFs) to get accurate results. This work considers only singles and doubles excitations with perturbative triple excitations through the RCI method.

The remaining part of the paper is organized as follows: In Sec. \ref{secth}, we present all the relevant mathematical forms of the operators used later in the calculations. In the next section, we briefly describe the MCDHF/RCI and RCC methods. Then, we discuss the results before concluding the work.

\begin{table}[t!]
    \centering
\caption{Layers of correlation orbitals, in non-relativistic notation, undertaken in the MCDHF/RCI method to carry out calculations of different atomic properties in Mg$^+$.}
    \begin{tabular}{ll} \hline \hline
  layers   &  orbitals\\ \hline
  layer 1       & $\{8s,8p,5d,4f\}$  \\
  layer 2       & $\{9s. 9p,  6d,5f,5g\}$ \\
  layer 3       & $\{10s,10p, 7d,6f,6g,6h\}$ \\
  layer 4       & $\{11s,11p, 8d,7f,7g,7h,7i\}$ \\
  layer 5       & $\{12s,12p, 9d,8f,8g,8h,8i\}$ \\
  layer 6       & $\{13s,13p,10d,9f,9g,9h,8i\}$\\
  layer 7       & $\{14s,14p,11d,10f,10g,10h,8i\}$\\ \hline \hline
    \end{tabular}
    \label{tab:layers}
\end{table}

\begin{table*}[t!]
\centering
\caption{Calculated excitation energies (in cm$^{-1}$) at different layers using the MCDHF/RCI method. The final results from the MCDHF/RCI method are taken from the ``layer 7" along with some corrections from the triple excitations (layer 7$+$T) and they are compared with the experimental values from the NIST ASD database \cite{NISTdata}. NCSFs is the total number of configuration state functions in the wave function expansion.
}
\begin{tabular}{c|c c c c c c c|c|r} \hline \hline
State & layer 1  & layer 2   & layer 3   & layer 4  & layer 5  & layer 6   & layer 7    & layer 7+T & Experiment  \\ \hline
       $3s~^2S_{1/2}$     &      0    &      0    &      0    &      0    &      0    &        0  &        0  &        0 &      0       \\ 
       $3p~^2P^o_{1/2}$   & 35742.44~ & 35648.53~ & 35555.39~ & 35547.24~ & 35545.12~ & 35544.12~ & 35544.13~ & 35576.67 & 35669.31  \\
       $3p~^2P^o_{3/2}$   & 35839.13~ & 35739.69~ & 35646.15~ & 35637.97~ & 35635.92~ & 35634.91~ & 35635.04~ & 35667.68 & 35760.88  \\ 
       $4s~^2S_{1/2}$     & 69521.77~ & 69662.64~ & 69624.34~ & 69632.30~ & 69632.43~ & 69633.29~ & 69631.99~ & 69662.40 & 69804.95  \\
       $3d~^2D_{3/2}$     & 71139.14~ & 71324.45~ & 71291.88~ & 71303.64~ & 71307.31~ & 71307.54~ & 71308.06~ & 71346.15 & 71491.06  \\
       $3d~^2D_{5/2}$     & 71139.72~ & 71325.37~ & 71292.77~ & 71304.50~ & 71308.17~ & 71308.38~ & 71308.91~ & 71347.01 & 71490.19   \\ 
       $4p~^2P^o_{1/2}$   & 80381.37~ & 80482.75~ & 80416.93~ & 80422.22~ & 80422.21~ & 80423.02~ & 80421.92~ & 80460.44 & 80619.50  \\
       $4p~^2P^o_{3/2}$   & 80414.13~ & 80513.16~ & 80447.19~ & 80452.48~ & 80452.49~ & 80453.30~ & 80452.24~ & 80490.79 & 80650.02  \\ \hline
       NCSFs              & 313875    &  633060   & 1141855   & 1887910   & 2852859   & 3912274   &  5141775  & 25605619 &       \\
       \hline \hline
    \end{tabular}
    \label{tab_exeCI}
\end{table*}

\begin{table*}[t]
\centering
\caption{Calculated second ionization potential and excitation energies (in cm$^{-1}$) at different levels of approximation in the RCC theory and comparison with the experimental values \cite{NISTdata}.}
\begin{tabular}{c|ccc c cc| cc} \hline \hline
State  & DHF  &  RCCSD  & RCCSDT & Basis  &  Breit & QED & Final & Experiment  \\ \hline
\multicolumn{9}{c}{Second ionization potential} \\        
$3s~^2S_{1/2}$  &  118823.96  &  121182.36  & 121250.59  & 14.15 & $-8.13$  & $-8.09$ & 121249(15)  & 121267.64  \\
\hline \\ 
\multicolumn{9}{c}{Excitation energies} \\ 
$3p~^2P^o_{1/2}$   & 34530.15    &    35631.66  &  35659.64  & 4.99 & $0.53$ & $-8.91$ &  35656(20)  & 35669.31 \\
$3p~^2P^o_{3/2}$   & 34620.49    &    35729.20  &  35757.28  & 5.05 & $-4.85$ & $-8.65$ & 35749(20)   & 35760.88  \\
$4s~^2S_{1/2}$     & 67967.16    &    69739.45  &  69792.55  & 10.68 &  $-5.83$ & $-5.89$ & 69792(18)   & 69804.95  \\
$3d~^2D_{3/2}$     & 69482.97  &  71427.61  &  71485.35  & 6.73 & $-8.60$ & $-8.09$ &  71475(18)  & 71491.06  \\
$3d~^2D_{5/2}$     &  69482.14   & 71426.88  &  71484.59  &  6.74 &  $-8.61$ & $-8.08$ & 71475(18)   & 71490.19  \\
$4p~^2P^o_{1/2}$   &  78576.27  &  80549.97  &  80606.04  & 11.27 & $-5.16$ & $-8.37$ & 80604(16)  & 80619.50 \\
$4p~^2P^o_{3/2}$   &  78606.96  &  80582.51  &  80638.62  &  11.28 & $-6.99$ & $-8.28$ & 80635(16) &  80650.02 \\
\hline \hline
\end{tabular}
\label{tab_execc}
\end{table*}

\section{Theory} \label{secth}

It is necessary to evaluate the single particle matrix elements to determine either expectation values or matrix elements of an operator using atomic wave functions. In the relativistic framework, the single particle electron wave function $|\phi \rangle$ is given by
\begin{eqnarray}
|\phi \rangle &=& \frac{1}{r} \begin{pmatrix} 
     P(r) & \chi_{\kappa, m_j}(\theta, \phi) \\
     \iota Q(r) & \chi_{-\kappa, m_j}(\theta, \phi) 
   \end{pmatrix}  ,
\end{eqnarray}
where $P(r)$ and $Q(r)$ are the large and small components of the radial part of the wave function, and $\chi_{\kappa, m_j}(\theta, \phi)$ is the angular factor with the relativistic angular momentum quantum number $\kappa$ and azimuthal component $m_j$ of the total angular momentum $j$. 

The single particle matrix element of the E1 operator ($D=\sum_q d_q$) in the length gauge is given by  
\begin{eqnarray}
\langle \phi_f | d_q | \phi_i \rangle = \delta ( (-1)^{l_f} , (-1)^{l_i+1}) (-1)^{j_f+m_f} \begin{pmatrix} 
     j_f & 1 & j_i \\
     -m_f & q & m_i 
   \end{pmatrix}   \nonumber \\ 
  \times (-1)^{j_f+1/2}  \sqrt{(2j_f+1)(2j_i+1)} \begin{pmatrix} 
     j_f & 1 & j_i \\
     \frac{1}{2} & 0 & -\frac{1}{2} 
   \end{pmatrix}  \nonumber \\
   \times \int_0^{\infty} \ dr \ r \left \{ (P_f(r)P_i(r)+Q_f(r)Q_i(r)) \right. \nonumber \\
   \left. - \frac{(\epsilon_f - \epsilon_i) }{ 5 \alpha } r
   \left [ \frac{(\kappa_f-\kappa_i )}{2} (P_f(r)Q_i(r)+Q_f(r)P_i(r)) \right. \right. \nonumber \\  \left. \left. +  (P_f(r)Q_i(r)-Q_f(r)P_i(r)) \right ] 
   \right \}, \ \ \ \ \ 
\end{eqnarray} 
where $\epsilon$s are the single particle orbital energies.

The transition probability (in s$^{-1}$) of an atomic state ($|\Psi_i \rangle$) to a decay state ($|\Psi_f \rangle$) due to the E1 decay channel is given by 
\begin{eqnarray}
A_{i \rightarrow f} = \frac{2.02613 \times 10^{18}}{(2J_i +1) \lambda_{i \rightarrow f}^3} |\langle J_i || D || J_f \rangle|^2 ,
\end{eqnarray}
where $\lambda_{i \rightarrow f}$ is the transition wavelength (in \AA) and $\langle J_i || D || J_f \rangle$  the reduced E1 transition matrix element (in atomic units (a.u.)) with the angular momentum of the states $J$s. An electron can decay from $|\Psi_i \rangle$ to all possible lower states. Thus, the branching ratio ($\Gamma_{i \rightarrow}$) of the decay from the state $|\Psi_i \rangle$ to the state $|\Psi_f \rangle$ is given by
\begin{eqnarray}
\Gamma_{i \rightarrow} = \frac{A_{i \rightarrow f}} {\sum\limits_f A_{i \rightarrow f} } .
\end{eqnarray}

The lifetime (in s) of the state $|\Psi_i \rangle$ can be determined by 
\begin{eqnarray}
\tau_i = \frac{1} {\sum\limits_f A_{i \rightarrow f} } .
\end{eqnarray}

Similarly, the static polarizability ($\alpha_d$) of the ground state of Mg$^+$ can be evaluated by the expression
\begin{eqnarray}
\alpha_d = - \frac{2}{3(2J_0 +1)} \sum_{n \ne 0} \frac{|\langle J_0 || D || J_n \rangle|^2}{E_0 - E_n} ,  
\end{eqnarray}
where $J_0$ and $J_n$ are the angular momenta of the ground state and excited state ($n$) with the respective energies $E_0$ and $E_n$.

The expression for $A_{hf}$ is given by
\begin{eqnarray}
A_{hf} = \mu_N g_I \frac{\langle J || T_{hf}^{(1)}|| J\rangle}{\sqrt{J(J+1)(2J+1)}} ,
\end{eqnarray}
where $\mu_N$ is the nuclear magneton and $T_{hf}^{(1)}$ is magnetic dipole hyperfine structure operator
\begin{eqnarray}\label{eq:Aop}
T_{hf}^{(1)} = \sum_k^{N_e} t_q^{(1)}(r_k) \equiv \sum_k - \iota e \sqrt{\frac{8 \pi}{3}} \frac{{\vec \alpha} \cdot {\vec Y}^1_q({\hat r}_k)}{r_k^2}.
\end{eqnarray}
For the latter, the sum is over the number of electrons $N_e$. The single particle matrix element of $t_q^{(1)}$ is given by
\begin{eqnarray}
\langle \phi_f | t_q^{(1)} | \phi_i \rangle &=& \delta ( (-1)^{l_f} , (-1)^{l_i})  (\kappa_f + \kappa_i) \nonumber \\ && (-1)^{j_f+m_f} \begin{pmatrix} 
     j_f & 1 & j_i \\
     -m_f & q & m_i 
   \end{pmatrix}  (-1)^{j_f+1/2} \nonumber \\ 
  \times &&  \sqrt{(2j_f+1)(2j_i+1)} \begin{pmatrix} 
     j_f & 1 & j_i \\
     \frac{1}{2} & 0 & -\frac{1}{2} 
   \end{pmatrix}  \nonumber \\
   \times &&  \int_0^{\infty} \ dr \ \frac{(P_f(r)Q_i(r)+Q_f(r)P_i(r))}{r^2} ,
\end{eqnarray} 
where $l_i$ is orbital angular momentum of the $i^{th}$ orbital.

The first-order IS $\delta E^{A,A'}$ of an atomic state between isotopes $A$ and $A'$ can be approximated as
\begin{equation}
\delta E^{A,A'} = F~\delta \langle r_N^2 \rangle^{A,A'} + K^{\text{MS}}(\mu_{A'}-\mu_A), 
\label{IS}
\end{equation}
where $F$ and $K^{\text{MS}}$ are known as the field-shift (FS) and mass-shift (MS) factors, respectively, $\delta \langle r_N^2 \rangle^{A,A'}\equiv\langle r_N^2 \rangle^{A'}-\langle r_N^2 \rangle^{A}$ is the change in the mean square nuclear radius between the two isotopes and $\mu_A=(M_A+m_e)^{-1}$ with $m_e$ is the electron mass and $M_A$ is the nuclear mass. The MS constant is further written as
\begin{eqnarray}
 K^{MS}  = K^{NMS} + K^{SMS} , 
\end{eqnarray}
where $K^{NMS}$ and $K^{SMS}$ are known as the normal mass shift (NMS) and specific mass shift (SMS) constants, respectively. The FS, NMS, and SMS constants can be obtained by calculating the expectation values of the corresponding operators. 

The FS operator is given as
\begin{eqnarray}\label{eq:Fop}
\bar{F} = \sum_i^{N_e} f (r_i) = - \sum_i^{N_e} \frac{\delta V_n(r_i)^{A',A}}{\delta \langle r_N^2 \rangle^{A',A} } ,
\end{eqnarray}
where $\delta V_n(r_i)^{A,A'}$ is the change in the nuclear potential ($V_n(r)$) between isotopes $A$ and $A'$. The FS factor $F$ can be calculated from its respective operator $\bar{F}$ using atomic theory methods. The single particle matrix element of the FS operator is given by
\begin{eqnarray}
\langle \phi_f | f | \phi_i \rangle 
&=&  \delta(\kappa_f,\kappa_i) \delta(m_{f}, m_{i}) \times \nonumber \\
&&\int_0^{\infty} dr  f(r) \left ( P_f(r) P_i (r) + Q_f(r) Q_i (r) \right ). \ \  
\end{eqnarray}  
In the RCC calculations, $F$ is defined by considering the fermi-charge distribution. However, we use $f = - \frac{2 \pi}{3} Z \rho_e(0)$ to estimate the FS constants using the MCDHF/RCI method \cite{EKMAN2019433} whose single particle matrix element is given by 
\begin{eqnarray}
\langle \phi_f | f | \phi_i \rangle 
&=&  - \delta(\kappa_f,\kappa_i) \delta(m_{f}, m_{i}) \frac{2 \pi}{3} \nonumber \\
&& \times \left ( P_f(0) P_i (0) + Q_f(0) Q_i (0) \right ), \ \  
\end{eqnarray}
So, it neglects variation of the electron density over the nuclear volume, which is a good approximation in the lighter Mg$^+$ system.

\begin{table*}[t!]
\centering
\caption{E1 transition rates (in $s$) of low-lying transitions in Mg$^+$ from the MCDHF/RCI method using length (shown as `l') and velocity (shown as `v') gauge expressions.}
\begin{tabular}{cc rrrrrrr} \hline \hline
State  & gauge & layer 1    & layer 2   & layer 3   & layer 4   & layer 5 & layer 6 & layer 7 \\ \hline
$3s~^2S_{1/2} \rightarrow 3p~^2P^o_{1/2}$  & l & ~2.623+08 & ~2.593+08 & ~2.576+08 & ~2.570+08 & ~2.571+08 & ~2.571+08 & ~2.571+08 \\
     & v &  2.577+08 & 2.571+08 & 2.562+08 & 2.564+08 & 2.565+08 & 2.565+08 & 2.565+08\\ [2ex]
     
 $3s~^2S_{1/2} \rightarrow 3p~^2P^o_{3/2}$  & l &  2.645+08 & 2.613+08 & 2.596+08 & 2.591+08 & 2.591+08 & 2.591+08 & 2.591+08 \\
         & v & 2.597+08 & 2.589+08 & 2.581+08 & 2.583+08 & 2.584+08 & 2.584+08 & 2.585+08 \\ [2ex]
         
$4s~^2S_{1/2} \rightarrow 3p~^2P^o_{1/2}$  & l & 1.145+08 & 1.148+08 & 1.148+08 & 1.151+08 & 1.151+08 & 1.151+08 & 1.150+08\\
     & v & 1.128+08 & 1.137+08 & 1.139+08 & 1.145+08 & 1.145+08 & 1.145+08 & 1.145+08\\ [2ex]
     
$4s~^2S_{1/2} \rightarrow 3p~^2P^o_{3/2}$  & l & 2.289+08 & 2.294+08 & 2.294+08 & 2.300+08 & 2.299+08 & 2.299+08 & 2.299+08 \\
      & v & 2.254+08 & 2.271+08 & 2.276+08 & 2.289+08 & 2.288+08 & 2.288+08 & 2.287+08\\ [2ex]
      
$3s~^2S_{1/2} \rightarrow 4p~^2P^o_{1/2}$  & l & 7.416+05 & 1.026+06 & 1.151+06 & 1.313+06 & 1.304+06 & 1.296+06 & 1.288+06\\
   & v & 1.141+06 & 1.275+06 & 1.310+06 & 1.310+06 & 1.299+06 & 1.300+06 & 1.294+06\\ [2ex]
   
$3s~^2S_{1/2} \rightarrow 4p~^2P^o_{3/2}$  & l & 6.191+05 & 8.900+05 & 1.006+06 & 1.158+06 & 1.148+06 & 1.141+06 & 1.134+06\\
        & v & 9.973+05 & 1.132+06 & 1.162+06 & 1.161+06 & 1.151+06 & 1.151+06 & 1.145+06\\ [2ex]
        
$4s~^2S_{1/2} \rightarrow 4p~^2P^o_{1/2}$  & l & 3.682+07 & 3.641+07 & 3.616+07 & 3.606+07 & 3.606+07 & 3.606+07 & 3.607+07\\
         & v & 3.622+07 & 3.588+07 & 3.572+07 & 3.611+07 & 3.584+07 & 3.583+07 & 3.583+07\\ [2ex]
         
$4s~^2S_{1/2} \rightarrow 4p~^2P^o_{3/2}$  & l & 3.712+07 & 3.669+07 & 3.644+07 & 3.634+07 & 3.634+07 & 3.634+07 & 3.635+07\\
     & v & 3.651+07 & 3.615+07 & 3.599+07 & 3.611+07 & 3.611+07 & 3.611+07 & 3.610+07\\ [2ex]
     
$3p~^2P^o_{1/2} \rightarrow 3d~^2D_{3/2}$  & l & 3.946+08 & 4.001+08 & 4.016+08 & 4.019+08 & 4.021+08 & 4.023+08 & 4.023+08\\
              & v & 3.928+08 & 4.003+08 & 4.021+08 & 4.030+08 & 4.034+08 & 4.037+08 & 4.038+08\\ [2ex]
              
$3p~^2P^o_{3/2} \rightarrow 3d~^2D_{3/2}$  & l & 7.852+07 & 7.964+07 & 7.995+07 & 8.000+07 & 8.005+07 & 8.007+07 & 8.008+07\\
            & v & 7.819+07 & 7.969+07 & 8.005+07 & 8.022+07 & 8.030+07 & 8.037+07 & 8.039+07\\ [2ex]
            
$3p~^2P^o_{3/2} \rightarrow 3d~^2D_{5/2}$  & l & 4.711+08 & 4.778+08 & 4.796+08 & 4.800+08 & 4.802+08 & 4.804+08 & 4.805+08\\
         & v & 4.689+08 & 4.780+08 & 4.802+08 & 4.812+08 & 4.817+08 & 4.821+08 & 4.822+08\\ [2ex]
         
 $4p~^2P^o_{1/2} \rightarrow 3d~^2D_{3/2}$  & l &  1.711+07 & 1.672+07 & 1.657+07 & 1.654+07 & 1.652+07 & 1.650+07 & 1.650+07\\
         & v & 1.669+07 & 1.639+07 & 1.630+07 & 1.635+07 & 1.636+07 & 1.642+07 & 1.641+07\\ [2ex]
         
$4p~^2P^o_{3/2} \rightarrow 3d~^2D_{3/2}$  & l & 1.725+06 & 1.684+06 & 1.669+06 & 1.666+06 & 1.664+06 & 1.662+06 & 1.662+06\\
          & v & 1.684+06 & 1.652+06 & 1.643+06 & 1.648+06 & 1.649+06 & 1.655+06 & 1.654+06 \\ [2ex]
          
$4p~^2P^o_{3/2} \rightarrow 3d~^2D_{5/2}$  & l &  1.552+07 & 1.516+07 & 1.502+07 & 1.499+07 & 1.497+07 & 1.496+07 & 1.496+07 \\
   & v &  1.515+07 & 1.487+07 & 1.479+07 & 1.484+07 & 1.484+07 & 1.490+07 & 1.489+07\\ 
\hline \hline
\end{tabular}
\label{tab_A}
\end{table*}

\begin{table*}[t!]
\centering
\caption{The estimated E1 matrix elements (in a.u.) of low-lying transitions in Mg$^+$ from the MCDHF/RCI method using the length gauge expression.}
\begin{tabular}{c|rrrrrrr} \hline \hline
Transition   & layer 1  & layer 2 & layer 3 & layer 4 & layer 5 & layer 6   & layer 7 \\ \hline
$3s~^2S_{1/2} \rightarrow 3p~^2P^o_{1/2}$  &  ~2.381  & ~2.377 & ~2.378 & ~2.377 & ~2.377 & ~2.377  & ~2.377\\
$3s~^2S_{1/2} \rightarrow 3p~^2P^o_{3/2}$  &  3.368  & 3.362 & 3.364 & 3.362 & 3.362 & 3.362 & 3.362\\
$4s~^2S_{1/2} \rightarrow 3p~^2P^o_{1/2}$  &  1.713  & 1.697 & 1.693 & 1.694 & 1.693 & 1.693 & 1.693\\
$4s~^2S_{1/2} \rightarrow 3p~^2P^o_{3/2}$  &  2.431  & 2.408 & 2.402 & 2.404 & 2.403 & 2.403 & 2.403\\
$3s~^2S_{1/2} \rightarrow 4p~^2P^o_{1/2}$  &  ~0.0375 & ~0.0440& ~0.0467& ~0.0499& ~0.0497& ~0.0496 & 0.0495\\
$3s~^2S_{1/2} \rightarrow 4p~^2P^o_{3/2}$  &  0.0484 & 0.0580& 0.0617& 0.0663& 0.0660& 0.0658 & 0.0656 \\
$4s~^2S_{1/2} \rightarrow 4p~^2P^o_{1/2}$  &  5.327  & 5.326 & 5.328 & 5.323 & 5.323 & 5.324 & 5.324\\
$4s~^2S_{1/2} \rightarrow 4p~^2P^o_{3/2}$  &  7.531  & 7.530 & 7.533 & 7.525 & 7.526 & 7.526 & 7.526\\
$3p~^2P^o_{1/2} \rightarrow 3d~^2D_{3/2}$  &  4.191  & 4.170 & 4.168 & 4.166 & 4.166 & 4.166 & 4.167\\
$3p~^2P^o_{3/2} \rightarrow 3d~^2D_{3/2}$  &  1.877  & 1.867 & 1.866 & 1.866 & 1.865 & 1.866 & 1.866\\
$3p~^2P^o_{3/2} \rightarrow 3d~^2D_{5/2}$  &  5.631  & 5.603 & 5.600 & 5.597 & 5.597 & 5.598 & 5.598\\
$4p~^2P^o_{1/2} \rightarrow 3d~^2D_{3/2}$  &  4.626  & 4.636 & 4.641 & 4.641 & 4.641 & 4.639 & 4.639\\
$4p~^2P^o_{3/2} \rightarrow 3d~^2D_{3/2}$  &  2.066  & 2.070 & 2.073 & 2.073 & 2.073 & 2.072 & 2.072\\
$4p~^2P^o_{3/2} \rightarrow 3d~^2D_{5/2}$  &  6.197  & 6.211 & 6.217 & 6.218 & 6.218 & 6.215 & 6.215\\
\hline \hline 
\end{tabular}
\label{tab_matE1}
\end{table*}

The relativistic forms of the NMS and SMS operators, up to the order of $\alpha Z$, in atomic units (a.u.) are given by
\begin{eqnarray}\label{nmsexp}
O^{\text{NMS}} &=& \sum_i^{N_e} o_i^\text{NMS} \nonumber \\
&\equiv&\frac{1}{2}\sum_i^{N_e} \left ({\vec p}_i^{~2} - \frac{\alpha Z}{r_i} {\vec \alpha}_i^D \cdot {\vec p}_i 
- \frac{\alpha Z}{r_i} ({\vec \alpha}_i^D \cdot {\vec C}_i^1){\vec C}_i^1 \cdot {\vec p}_i \right ) \nonumber 
\end{eqnarray}
and 
\begin{eqnarray}\label{smsexp}
\hspace{-13mm}
O^{\text{SMS}} = \sum_{i \ne j}^{N_e} o_{ij}^\text{SMS} 
&\equiv& \frac{1}{2} \sum_{i\ne j}^{N_e} \left ({\vec p}_i \cdot {\vec p}_j - \frac{\alpha Z}{r_i} {\vec \alpha}_i^D \cdot {\vec p}_j 
 \right. \nonumber \\ && \left. - \frac{\alpha Z}{r_i} ({\vec \alpha}_i^D \cdot {\vec C}_i^1) ({\vec p}_j \cdot {\vec C}_j^1) \right ), 
\end{eqnarray}
respectively, where ${\vec p}$ is the momentum operator, ${\vec \alpha}_i^D $ is the Dirac matrix, ${\vec C}^1$ is the Racah operator, and $\alpha$ is the fine-structure factor
\cite{Gaidamauskas_2011}.

The single particle matrix element of the NMS operator returns
\begin{eqnarray}
\langle \phi_f | o^\text{NMS} | \phi_i \rangle &=& \frac{1}{2} \delta(\kappa_f,\kappa_i) \delta(m_{f}, m_{i}) \nonumber \\ & \times & \int_0^{\infty} dr  \left ( \frac{\partial P_f(r)}{\partial r} \frac{\partial P_i(r)}{\partial r}  \right. \nonumber \\ 
&+& \left. \frac{\partial Q_f(r)}{\partial r} \frac{\partial Q_i(r)}{\partial r} +  \frac{l_i (l_i+1) P_f(r) P_i(r)} {r^2} \right. \nonumber \\ 
&+& \left. \frac{\tilde{l}_i (\tilde{l}_i+1) Q_f(r) Q_i(r) }{r^2} - 2 \frac{\alpha Z}{r} \right. \nonumber \\ &\times & \left. \left ( Q_f(r)\frac{\partial P_i(r)}{\partial r} + \frac{\partial P_f(r)}{\partial r} Q_i(r) \right ) -  \frac{\alpha Z}{r^2}  \right. \nonumber \\ &\times&  \left. (\kappa_i -1) \left ( Q_f(r) P_i(r) + P_f(r) Q_i(r) \right )  \right ) , \ \ \ \ \
\label{relnms}
\end{eqnarray} 
where $\tilde{l}$ is the orbital radial quantum numbers for the small component.
Similarly, the single particle matrix element of the SMS operator is given by
\begin{eqnarray}
\langle \phi_i \phi_j | o^\text{SMS} | \phi_k \phi_l \rangle &=& \delta(m_{i}-m_{k}, m_{l}-m_{j})  \nonumber \\ 
&& \times \sum_{q=-1}^{1} \begin{pmatrix} 
     j_i & 1 & j_k \\
     -m_i & q & m_k 
   \end{pmatrix} \begin{pmatrix} 
     j_j & 1 & j_l \\
     -m_j & -q & m_l 
   \end{pmatrix} \nonumber \\ 
&& \times (-1)^{j_i-m_i + j_j -m_j +1 -q} X^1(ij,kl) ,
\end{eqnarray}
where $X^1(ij,kl)$ is the reduced matrix element given by
\begin{eqnarray}
X^1(ij,kl) &=& \sqrt{(2j_i+1) (2j_j+1) (2j_k+1) (2j_l+1)} \nonumber \\
&& \times \begin{pmatrix} 
     j_i & 1 & j_k \\
     1/2 & 0 & -1/2 
   \end{pmatrix} \begin{pmatrix} 
     j_j & 1 & j_l \\
     1/2 & 0 & -1/2 
   \end{pmatrix} \nonumber \\
   && \times (-1)^{j_i+j_j+1} \left [ R(ik) R(jl) + \frac{1}{2} \left ( R(ik)X(jl) \right. \right. \nonumber \\ 
   && \left. \left. +X(ik)R(jl) \right ) \right ]
\label{relsms}
\end{eqnarray}
with the radial functions
\begin{eqnarray}
 R(ab) &=& - \iota \int_0^{\infty} dr \left [ P_a(r) \left ( \frac{\partial P_b(r)}{\partial r} \right. \right. \nonumber \\ &-& \left. \left. \frac{\kappa_a(\kappa_a-1)-\kappa_b(\kappa_b-1)}{2r} P_b(r) \right ) + Q_a(r) \left ( \frac{\partial Q_b(r)}{\partial r} \right. \right. \nonumber \\ &-& \left. \left.   \frac{\kappa_a(\kappa_a+1)-\kappa_b(\kappa_b+1)}{2r} Q_b(r) \right ) \right ] 
\end{eqnarray}
and
\begin{eqnarray}
 X(ab) &=& - \iota \int_0^{\infty} dr \frac{\alpha Z}{r} \left [ (\kappa_a - \kappa_b -2) P_a(r) Q_b(r) \right. \nonumber \\
 && \left. + (\kappa_a - \kappa_b  + 2) Q_a(r) P_b(r)\right ] . 
\end{eqnarray}

\begin{table*}[t!]
\centering
\caption{The calculated E1 matrix elements (in a.u.) at different levels of approximation in the RCC method using the length gauge expression.}
\begin{tabular}{c|ccc c c c | c} \hline \hline
Transition       &  DHF    &  RCCSD   & RCCSDT  &  Basis & Breit & QED & Final  \\ \hline
$3s~^2S_{1/2} \rightarrow 3p~^2P^o_{1/2}$  & 2.462 & 2.372 & 2.368 & $\sim 0.0$ & $\sim 0.0$ & $\sim 0.0$ & 2.368(2) \\
$3s~^2S_{1/2} \rightarrow 3p~^2P^o_{3/2}$  & 3.482 & 3.355 & 3.350 & $\sim 0.0$ & $\sim 0.0$ & 0.001 &  3.351(3) \\
$4s~^2S_{1/2} \rightarrow 3p~^2P^o_{1/2}$  &  1.706 & 1.695 & 1.694 & $\sim 0.0$ & $\sim 0.0$ & $\sim 0.0$ & 1.694(1)  \\
$4s~^2S_{1/2} \rightarrow 3p~^2P^o_{3/2}$  &  2.422 & 2.406 & 2.405 & $\sim 0.0$ & $\sim 0.0$ & $\sim 0.0$ & 2.405(1) \\
$3s~^2S_{1/2} \rightarrow 4p~^2P^o_{1/2}$  & 0.032 & 0.051 & 0.052 & $\sim 0.0$ & $\sim 0.0$ & $\sim 0.0$ & 0.052(1) \\
$3s~^2S_{1/2} \rightarrow 4p~^2P^o_{3/2}$  & 0.041 & 0.067 & 0.068 & $\sim 0.0$ & $\sim 0.0$ & $\sim 0.0$ & 0.068(1) \\
$4s~^2S_{1/2} \rightarrow 4p~^2P^o_{1/2}$  &  5.387 & 5.314 & 5.313 & $\sim 0.0$ & $-0.001$ & 0.001 & 5.313(1) \\
$4s~^2S_{1/2} \rightarrow 4p~^2P^o_{3/2}$  &  7.615 & 7.513 & 7.511 & $\sim 0.0$ & $\sim 0.0$ & 0.001 & 7.512(2) \\
$3p~^2P^o_{1/2} \rightarrow 3d~^2D_{3/2}$  & 4.268  & 4.162 & 4.158 & $\sim 0.0$ & 0.001 & $\sim 0.0$ & 4.159(2) \\
$3p~^2P^o_{3/2} \rightarrow 3d~^2D_{3/2}$  & 1.911  &  1.864 & 1.863 & $\sim 0.0$ & $\sim 0.0$ & $\sim 0.0$ & 1.863(1) \\
$3p~^2P^o_{3/2} \rightarrow 3d~^2D_{5/2}$  & 5.734  &  5.593 & 5.588 & $\sim 0.0$ & $\sim 0.0$ & $\sim 0.0$ & 5.588(3) \\
$4p~^2P^o_{1/2} \rightarrow 3d~^2D_{3/2}$  &  4.670 & 4.636 & 4.636 & $-0.001$ & $\sim 0.0$ & 0.001 & 4.636(1) \\
$4p~^2P^o_{3/2} \rightarrow 3d~^2D_{3/2}$  & 2.086  & 2.071 & 2.071 & $-0.001$ & $\sim 0.0$ & $\sim 0.0$ & 2.070(1) \\
$4p~^2P^o_{3/2} \rightarrow 3d~^2D_{5/2}$  & 6.257  & 6.212 &  6.211 & $-0.011$ & $-0.001$ & $\sim 0.0$ & 6.199(5)\\
\hline \hline
\end{tabular}
\label{tab_E1}
\end{table*}

\section{Methods and Approaches}

\subsection{The MCDHF/RCI method}

In the MCDHF method, the wave function of an atomic state labeled $\Gamma JM_J$ is expanded in terms of $jj$-coupled configuration state functions (CSFs)
\begin{equation}
    \Psi(\Gamma JM_J) = \sum_{\alpha=1}^{N_{\mbox{\scriptsize CSF}}} c^{\Gamma J}_{\alpha} \Phi(\gamma_{\alpha} JM_J),
\end{equation}
where $N_{\mbox{\scriptsize CSF}}$ denotes number of CSFs. The CSFs, $\Phi(\gamma_{\alpha} JM_J)$, are recursively constructed from products of one-electron Dirac orbitals, and the label $\gamma_{\alpha}$ contains information about the orbital occupancy and 
the coupling scheme \cite{fischer2016advanced}. Requiring the weighted energies of the targeted states, computed from the CSF expansions, to be
stationary with respect to perturbations in both the mixing coefficients and the radial orbitals leads to a matrix
eigenvalue problem
\begin{equation}
({\bm H} - E {\bm I}){\bm c} = {\bm 0},
\end{equation}
where ${\bm H}$ is the Hamiltonian matrix with elements between the CSFs based on the Dirac-Coulomb (DC) Hamiltonian operator, coupled with a set of radial integro-differential equations. 

In the MCDHF calculations, the eigenvalue problem and the integro-differential equations are solved iteratively in a self-consistent field procedure until 
convergence has been achieved \cite{grant2007relativistic, atoms11010007}.
In the RCI calculations, the radial functions are known, and only the eigenvalue problem is solved. Now, the Breit interaction and leading QED effects can be added to the Hamiltonian operator \cite{mackenzie1980program,grant2007relativistic}. The reduced Hamiltonian matrix elements between the CSFs,
$H_{\alpha \beta} = \langle  \Phi(\gamma_{\alpha} J) \|H\|  \Phi(\gamma_{\beta} J) \rangle$, needed for the matrix eigenvalue problem, but also for constructing the direct and exchange potentials
for the radial integro-differential equations, are obtained using spin-angular algebra based on angular momentum theory \cite{atoms10040129}. Introducing configuration state function generators  (CSFGs) as described in \cite{YTLiCPC}, the spin-angular integrations can be kept at a minimum, substantially reducing the time for the calculations.

\begin{table}[t!]
\centering
\caption{The estimated lifetimes (in $ns$) of low-lying states of Mg$^+$ by combining the E1 matrix elements from the MCDHF/RCI and RCC methods with the experimental wavelengths (derived from the experimental energies \cite{NISTdata}). The estimated branching ratios (BR) are also given. Our RCC values, with the estimated uncertainties, are compared with the available measurements.}
\begin{tabular}{cc c ccc} \hline \hline
Atomic & Decay    &   BR  & \multicolumn{3}{c}{Lifetime (in ns)}  \\ 
\cline{4-6} \\
 State  & State & $\Gamma$  & \footnotesize{MCDHF} & \footnotesize{RCC} & \footnotesize{Experiment} \\ 
 $|\Psi_i \rangle$ & $|\Psi_f \rangle$ &   & \footnotesize{/RCI} &  &  \\
\hline \\
$3p~^2P^o_{1/2}$ & $\rightarrow 3s~^2S_{1/2}$ & 1.0  & 3.889 & 3.879(7) & 3.854(30) \cite{Ansbacher1989} \\
                 &                            &       &      &     & 4.0(3) \cite{Lundin1973} \\
                 &                            &       &      &     & 6.20(38) \cite{Schaefer1971} \\
              &                            &       &      &     & 4.2(4) \cite{Andersen1970} \\
              &                            &       &      &     & 4.5(8) \cite{Berry1970} \\                
$3p~^2P^o_{3/2}$ & $\rightarrow 3s~^2S_{1/2}$ & 1.0  & 3.859 & 3.844(7) & 3.810(40)  \cite{Ansbacher1989} \\
                 &                            &       &      &     & 4.0(3) \cite{Lundin1973} \\
                 &                            &       &      &     & 6.20(38) \cite{Schaefer1971} \\   
              &                            &       &      &     & 4.2(4) \cite{Andersen1970} \\
               &                            &       &      &     & 4.5(8) \cite{Berry1970} \\                
$4s~^2S_{1/2}$ & $\rightarrow 3p~^2P^o_{3/2}$ & 0.67 & 2.898 & 2.883(2) & 2.6(3) \cite{Lundin1973} \\
              &                            &       &      &     & 2.8(4) \cite{Andersen1970} \\
& $\rightarrow 3p~^2P^o_{1/2}$ &  0.33 &  &  &  3.8(5) \cite{Berry1970} \\
$3d~^2D_{5/2}$ & $\rightarrow 3p~^2P^o_{3/2}$ & 1.0 & 2.081 & 2.079(2) & 2.2(2)  \cite{Lundin1973} \\
              &                            &       &      &     & 1.9(2) \cite{Andersen1970} \\
              &                            &       &      &     & 2.3(4) \cite{Berry1970} \\
$3d~^2D_{3/2}$ & $\rightarrow 3p~^2P^o_{3/2}$ &  0.17 & 2.073 & 2.071(2) & 2.2(2) \cite{Lundin1973} \\ 
              &                            &       &      &     & 1.9(2) \cite{Andersen1970} \\
              &                            &       &      &     & 2.3(4) \cite{Berry1970} \\
& $\rightarrow 3p~^2P^o_{1/2}$ &  0.83 &  & \\ 
$4p~^2P^o_{1/2}$  & $\rightarrow 3d~^2D_{3/2}$  & 0.30  & 18.566  & 18.462(26) & 21(2) \cite{Lundin1973} \\
  & $\rightarrow 4s~^2S_{1/2}$ & 0.67 &  & & \\
    & $\rightarrow 3s~^2S_{1/2}$ & 0.03 &  &  & \\
$4p~^2P^o_{3/2}$ & $\rightarrow 3d~^2D_{3/2}$  & 0.03  & 18.480 & 18.411(27) &  21(2) \cite{Lundin1973}  \\
 &  $\rightarrow 3d~^2D_{5/2}$  &  0.28 &  &  & \\
& $\rightarrow 4s~^2S_{1/2}$ &  0.67 &  & & \\
& $\rightarrow 3s~^2S_{1/2}$ & 0.02 & &  & \\
\hline \hline
\end{tabular}
\label{tab_LMCDHF}
\end{table}


Once the wave functions of the states have been determined, 
properties are evaluated in terms of reduced matrix elements of the corresponding electronic operators
\begin{equation}
\langle \Gamma J \|O\| \Gamma J \rangle = \sum_{\alpha \beta}^{N_{\mbox{\scriptsize CSF}}} c^{\Gamma J}_{\alpha} c^{\Gamma J}_{\beta}  \langle \Phi(\gamma_{\alpha} J)\| O\| \Phi(\gamma_{\beta} J) \rangle.
\end{equation}
Spin-angular integration resolves the matrix elements between
CSFs in spin-angular coefficients and radial integrals which are multiplied and accumulated.
The matrix elements are then weighted by the product of the expansions coefficients \cite{atoms11010007}. Details can be found in \cite{LI2020107211} for hyperfine structure and in \cite{EKMAN2019433} for the evaluation of the isotope shift parameters.
Transition parameters for E1 transitions are given in terms of the squared transition amplitude
\begin{equation}
\langle \Gamma J \| D \| \Gamma' J' \rangle = \sum_{\alpha \beta}^{N_{\mbox{\scriptsize CSF}}} c^{\Gamma J}_{\alpha} c^{\Gamma' J'}_{\beta}  \langle \Phi(\gamma_{\alpha} J\| D \| \Phi(\gamma_{\beta} J') \rangle.
\end{equation}
The matrix elements between the CSFs, in turn, are given as one-electron matrix elements
$\langle n_a \kappa_a \| d \| n_b \kappa_b \rangle $ weighted with spin-angular coefficients. Following Grant \cite{Grant1974,DYALL1989425} we have
\begin{eqnarray}
    \langle n_a \kappa_a \| D \| n_b \kappa_b \rangle  =   
    \left( \frac{(2j_b+1) \omega}{\pi c} \right)^{1/2} \times  \nonumber \\  (-1)^{j_a-1/2} 
    \left(
    \begin{array}{ccc}
    j_a & 1 & j_b \\
    1/2 & 0 & -1/2 \\
    \end{array}\right)
    (\overline{M}^e_{ab} + G \overline{M}^l_{ab}),
\end{eqnarray}
where $\omega$ is the transition frequency and $\overline{M}^e_{ab}$ and $\overline{M}^l_{ab}$ are radiative transition integrals. $G$ is the gauge parameter. The necessary and sufficient condition for the transition amplitudes for all multipoles to be gauge invariant is that $\overline{M}^l_{ab}$ should vanish identically. For approximate calculations, this is not the case, and instead we have a parabolic behavior in $G$ for the transition parameters. Traditionally, the transition parameters have been given in the Coulomb gauge  with $G=0$ (reduces to the velocity form in the non-relativistic limit) and the Babushkin gauge with $G = \sqrt{2}$ (reduces to the length form in the non-relativistic limit). The agreement between transitions parameters in the above two gauges has for a long time incorrectly been used to estimate the uncertainty of the result \cite{ekman2014validation}. This is due to the fact, in general speaking, that the dependence of the transition parameter on $G$ is parabolic~\cite{Gaigalas.2020.p13} and that transition parameters in Babushkin and Coulomb gauges can be on different branches of it~\cite{Rynkun.2022.p82}. The agreement between these two gauges can thus be accidental, and the closeness of the two gauges does not indicate the accuracy. Therefore, this method of estimating the uncertainty of the results cannot be used as an accuracy indicator. A new method, the quantitative and qualitative evaluation (QQE) approach developed by Gaigalas and coworkers \cite{Gaigalas.2022.p281,Kitoviene.2024.p}, does not suffer from the above disadvantage. Instead it evaluates the transition parameters over the full range of the gauge parameter and also includes an analysis of the relative signs of the Coulomb and Babushkin contributions. The quantitative and qualitative approach is more accurate and makes it possible to estimate the uncertainty of the transition rates according to the National Institute of Science and Technology Atomic Spectra Atomic Spectroscopic Data (NIST ASD)~\citep{Kramida.2023.p} terminology (AA $\leq$ 1~\%, A$^{+}$ $\leq$ 2~\%, A $\leq$ 3~\%, B$^{+}$ $\leq$ 7~\%, B $\leq$ 10~\%, C$^{+}$  $\leq$ 18~\%,  C $\leq$ 25~\%,  D$^{+}$ $\leq$ 40~\%, D $\leq$ 50~\%, and E $>$ 50~\% ).

\subsection{The RCC method}

In the RCC theory {\it ans\"atz}, the wave function of an atomic state is defined as \cite{cizek}
\begin{eqnarray}
|\Psi_0 \rangle = e^T |\Phi_0 \rangle .
\end{eqnarray}
In the present work, $|\Psi_0 \rangle$ is the ground state wave function of the $[2p^6]$ closed-core of Mg$^+$ and $|\Phi_0 \rangle$ is its DHF wave function. Note that we start calculations with this closed-core so that all the interested excited states of Mg$^+$, which have an extra valence electron over this closed-core, can be produced by appending the respective valence orbital in the Fock-space approach. Thus, an atomic state of Mg$^+$ with the closed-core $[2p^6]$ and a valence orbital $v$ is expressed in the RCC theory as \cite{mukherjee84,lindgren87}
\begin{eqnarray}
|\Psi_v \rangle = e^T \left \{ 1+S_v \right \}|\Phi_v \rangle ,
\end{eqnarray}
where $|\Phi_v \rangle = a_v^{\dagger} |\Phi_0 \rangle$ and $S_v$ is a new RCC operator that takes into correlation among valence electron with other core electrons.

The amplitude-determining equation for the $T$ operator is given by \cite{bartlett}
\begin{eqnarray}
\langle \Phi_0^* | \bar{H}  | \Phi_0 \rangle = 0 ,
\label{eqs0}
\end{eqnarray}
where $| \Phi_0^* \rangle $ represents all possible excited state determinants with respect to $| \Phi_0 \rangle$ and $\bar{H}= e^{-T} H e^T = (He^T)_l$ with the subscript $l$ denoting the linked terms. 

\begin{table}[t!]
\centering
\caption{Contributions from different E1 matrix elements from the MCDHF/RCI and RCC methods to the ground state electric dipole polarizability (in a.u.) of Mg$^+$. The tail, core-valence and core contributions are estimated using the RPA. The final values from both the MCDHF/RCI and RCC methods are compared with the values previously obtained using other methods and experiment.}
\begin{tabular}{l l c c} \hline \hline
 Reference   & Source   &  MCDHF/RCI & RCC \\ 
\hline \\
This work &   &   \\
    & $3p~^2P^o_{1/2}$ & 11.59  & 11.50(2) \\
    & $3p~^2P^o_{3/2}$ & 23.12  & 22.97(4) \\
    & $4p~^2P^o_{1/2}$ & $\sim 0.0$ &  $\sim 0.0(1)$ \\
    & $4p~^2P^o_{3/2}$  & 0.01 & 0.01(1) \\
 & Tail    &  \multicolumn{2}{c}{0.02(2)} \\
 & Core-valence    & \multicolumn{2}{c}{$-0.02(2)$} \\
 & Core  &   \multicolumn{2}{c}{0.44(4)} \\ 
\hline \\
   & Final &  35.16   &  34.92(7)  \\
 Experiment &  &  \multicolumn{2}{c}{$33.80^{+0.50}_{-0.30}$} \cite{Mgionpol} \\
\hline \hline
\end{tabular}
\label{tab_pol}
\end{table}

\begin{table*}[t!]
\centering
\caption{The calculated $A_{hf}$ values (in MHz) at different layers using the MCDHF/RCI method. The final results from the MCDHF/RCI method are taken from the ``layer 7" along with some corrections from the triple excitations (layer 7$+$T), and they are compared with the experimental values.}
\begin{tabular}{c|rrrrrrr|r|r} \hline \hline
State  & layer 1    & layer 2    & layer 3    & layer 4    & layer 5    & layer 6   & layer 7    & layer 7+T & Experiment \\ \hline
$3s~^2S_{1/2}$     & $-556.71$ & $-567.85$ & $-575.60$ & $-576.21$ & $-589.28$ & $-589.40$ & $-589.82$ & $-592.48$ & $-596.2542487(42)$ \cite{Xu2017} \\ 
               &     &    &   &  &  &  &  &  & $-596.2544(5)$ \cite{Itano1981} \\
$3p~^2P^o_{1/2}$   &  $-97.68$ &  $-98.44$ & $-100.17$ & $-100.60$ & $-101.11$ & $-101.12$ & $-101.36$ & $-101.81$ &  \\  
$3p~^2P^o_{3/2}$   &  $-20.61$ &  $-20.71$ & $-19.26$  & $-19.21$  & $-18.69$  & $-18.69$  & $-18.79$  & $-18.97$  &   \\ 
$4s~^2S_{1/2}$     & $-150.82$ & $-153.51$ & $-156.78$ & $-157.62$ & $-161.21$ & $-161.26$ & $-161.36$ & $-162.16$ &  \\
$3d~^2D_{3/2}$     &   $-1.19$ &  $-1.15$  & $-1.14$   & $-1.14$   & $-1.13$   & $-1.14$   & $-1.14$   & $-1.13$   &  \\
$3d~^2D_{5/2}$     &    0.17   &  0.071    & $0.055$   & $0.046$   & $0.032$   & $0.033$   & $0.035$   & 0.071     &   \\ 
$4p~^2P^o_{1/2}$   &  $-32.32$ &  $-32.37$ & $-33.02$  & $-33.35$  & $-33.55$  & $-33.55$  & $-33.63$  & $-33.80$  &   \\
$4p~^2P^o_{3/2}$   &   $-6.68$ &  $-6.76$  & $-6.27$   & $-6.30$   & $-6.13$   & $-6.13$   & $-6.16$   & $-6.22$   &  \\ 
\hline \hline
\end{tabular}
\label{tab_hfs_MCDHF}
\end{table*}

\begin{table*}[t!]
\caption{The calculated $A_{hf}$ values (in MHz) of $^{25}$Mg$^+$ at different levels of approximation in the RCC theory and comparison with the experimental values. The corrections from Basis, BW, Breit, and QED are given from the RCCSD method. The final results include RCCSDT values with other corrections.}
\begin{tabular}{c|ccc c c c c | cc} \hline \hline
State       & DHF     &   RCCSD  & RCCSDT  &  Basis & BW & Breit & QED & Final &  Experiment \\ \hline
$3s~^2S_{1/2}$     & $-462.70$ &  $-595.70$ & $-597.97$ & $-0.27$ & 0.30 &  $-0.31$ & 1.63 & $-596.6(8)$  &   $-596.2542487(42)$ \cite{Xu2017} \\
               &     &    &   &  &  &  &  &  & $-596.2544(5)$ \cite{Itano1981} \\
$3p~^2P^o_{1/2}$   &  $-76.96$ & $-102.53$ &  $-103.31$ & $-0.03$ & $\sim 0.0$ & $-0.02$ & $\sim 0.0$ & $-103.4(5)$ &   \\
$3p~^2P^o_{3/2}$   &  $-15.24$ & $-19.12$ & $-19.29$  & $-0.01$ & $\sim 0.0$ & $-0.01$ & $\sim 0.0$ & $-19.31(5)$ & \\
$4s~^2S_{1/2}$     &  $-130.33$ & $-163.22$ &  $-163.61$  & $-0.06$ & 0.08 &  $-0.01$ &  0.42  & $-163.2(6)$ &  \\
$3d~^2D_{3/2}$     &   $-1.26$ & $-1.16$ &  $-1.15$ & $\sim 0.0$  & $\sim 0.0$ &  $\sim 0.0$ & $\sim 0.0$ & $-1.15(3)$ & \\
$3d~^2D_{5/2}$     &   $-0.54$  & $0.10$ & 0.14 & $\sim 0.0$  & $\sim 0.0$ &  0.01 & $\sim 0.0$ & 0.15(2) & \\
$4p~^2P^o_{1/2}$   &  $-26.15$ & $-34.09$  & $-34.40$ & $-0.01$  & $\sim 0.0$ & $-0.01$ & $\sim 0.0$ & $-34.42(8)$ & \\
$4p~^2P^o_{3/2}$   &  $-5.18$ & $-6.28$ &  $-6.32$ & $-0.01$ & $\sim 0.0$ & $\sim 0.0$ & $\sim 0.0$ & $-6.33(5)$ & \\
\hline \hline
\end{tabular}
\label{tab_hfs}
\end{table*}

The amplitude-determining equation for the $S_v$ operator is given by \cite{das}
\begin{eqnarray}
 \langle \Phi_v^* | \{ (\bar{H}-E_v) S_v \} + \bar{H} | \Phi_v \rangle = 0, \label{eqamp}
\end{eqnarray}
where $| \Phi_v^* \rangle$ is the excited state Slater determinant with respect to $| \Phi_v \rangle$. It should be worth noting that $E_v$ enters explicitly in Eq. (\ref{eqamp}), unlike calculations of the $T$ operator amplitudes. This is due to the fact that the DHF wave function used in Eq. (\ref{eqs0}) is obtained using the $V^{N_e}$ potential while it uses a $V^{N_e-1}$ potential in Eq. (\ref{eqamp}). Thus, the above equation is solved simultaneously with the energy determining equation, given by
\begin{eqnarray}
 E_v= \langle \Phi_v | \bar{H} \{ 1+ S_v \} | \Phi_v \rangle . \label{eqeng}
\end{eqnarray}

The last sub-section mentioned that the MCDHF/RCI method requires diagonalization of a square matrix of dimension $N$, with $N$ denoting NCSFs. In contrary, both Eqs. (\ref{eqs0}) and (\ref{eqamp}) of the RCC method utilize the Jacobi iterative scheme to solve the equation, which necessitates storing the matrix elements only in a single array of $N$ dimension. 

The E1 matrix elements and $A_{hf}$ values are evaluated in the RCC method as
\begin{eqnarray}
\langle O \rangle = \frac{\langle \Phi_f | \{1+S_f^{(0)} \}^{\dagger}  \bar{O} \{ 1+ S_i^{(0)} \} |\Phi_i \rangle} { \sqrt{ N_f N_i }} ,
\label{propeq}
\end{eqnarray}
where $O$ stands for one of the operator and $N_{v=f,i} = \langle \Phi_v | \{1+ S_v^{(0)} \}^{\dagger} \bar{N} \{ 1+ S_v^{(0)} \} |\Phi_v \rangle$. The above expression contains two non-terminating series, namely $\bar{O}=e^{T^{(0)\dagger}}Oe^{T^{(0)}}$ and $\bar{N}=e^{T^{(0)\dagger}}e^{T^{(0)}}$, which are solved by adopting a self-consistent approach \cite{dillip2014}. 

\begin{table*}[t!]
\centering
\caption{Calculated IS constants at different levels of the approximation in the MCDHF/RCI method.}
\begin{tabular}{c ccccccc r} \hline \hline 
State  & layer 1  & layer 2  & layer 3  & layer 4  & layer 5  & layer 6  & layer 7    & layer 7+T  \\ \hline
 & & & & & &  & \\
\multicolumn{9}{c}{FS constants MHz fm$^{-2}$ } \\
$3s~^2S_{1/2}$     & $68.27238$ &  $68.26874$ & $68.247105$ & $68.24102$ & $68.22743$ & 68.22733 & 68.24410 & $68.24436$  \\ 
$3p~^2P^o_{1/2}$   & $68.14787$ &  $68.14243$ &  68.120036  &  68.11406  &  68.10111  & 68.10101 & 68.11863 &  68.11898  \\  
$3p~^2P^o_{3/2}$   & $68.14783$ &  $68.14238$ &  68.120010  &  68.11403  &  68.10110  & 68.10099 & 68.11861 &  68.11896   \\ 
$4s~^2S_{1/2}$     & $68.18733$ &  $68.18266$ & $68.161376$ & $68.15549$ & $68.14231$ & 68.14221 & 68.15956 & $68.16027$  \\
$3d~^2D_{3/2}$     & $68.15586$ &  $68.15122$ &  68.129647  &  68.12367  &  68.11066  & 68.11056 & 68.12809 &  68.12885     \\
$3d~^2D_{5/2}$     & $68.15586$ &   68.15123  &  68.129649  &  68.12367  &  68.11067  & 68.11057 & 68.12810 &  68.12885    \\ 
$4p~^2P^o_{1/2}$   & $68.15323$ &  $68.14816$ &  68.126330  &  68.12033  &  68.10731  & 68.10721 & 68.12480 &  68.12546  \\
$4p~^2P^o_{3/2}$   & $68.15321$ &  $68.14814$ &  68.126320  &  68.12032  &  68.10731  & 68.10721 & 68.12479 &  68.12545\\ 
 & & & & & &  & \\
\multicolumn{9}{c}{NMS constants in GHz u} \\
$3s~^2S_{1/2}$     & $717778.69$ &  $719204.02$  & $720105.72$ & $720207.85$ & $720373.06$ & 720388.32 &  720634.30   &  720715.94   \\ 
$3p~^2P^o_{1/2}$   & $717226.06$ &  $718576.01$  &  719493.79  &  719595.22  &  719759.32  & 719774.34 &  720021.32   &  720119.75  \\  
$3p~^2P^o_{3/2}$   & $717224.42$ &  $718575.03$  &  719493.11  &  719594.49  &  719758.65  & 719773.67 &  720020.63   &  720119.17   \\ 
$4s~^2S_{1/2}$     & $716629.26$ &  $717980.67$  & $718905.50$ & $719007.04$ & $719168.31$ & 719183.38 &  719429.04   &  719554.02  \\
$3d~^2D_{3/2}$     & $716599.71$ &  $717940.02$  &  718866.32  &  718962.41  &  719123.60  & 719139.44 &  719385.06   &  719519.59  \\
$3d~^2D_{5/2}$     & $716599.85$ &  $717940.15$  &  718866.47  &  718962.56  &  719123.76  & 719139.59 &  719385.21   &  719519.74  \\ 
$4p~^2P^o_{1/2}$   & $716467.05$ &  $717798.85$  &  718724.44  &  718825.12  &  718986.49  & 719001.58 &  719247.65   &  719374.31   \\
$4p~^2P^o_{3/2}$   & $716466.59$ &  $717798.56$  &  718724.24  &  718824.87  &  718986.27  & 719001.36 &  719247.42   &  719374.11   \\ 
 & & & & & &  & \\
\multicolumn{9}{c}{SMS constants in GHz u} \\
$3s~^2S_{1/2}$     & $-92507.81$ & $-87879.83$ & $-83599.25$ & $-83302.79$ & $-80856.74$ & $-80859.30$ & $-80013.64$ & $-79959.12$  \\
$3p~^2P^o_{1/2}$   & $-92980.38$ & $-88181.67$ & $-83912.63$ & $-83609.13$ & $-81160.97$ & $-81161.58$ & $-80317.57$ & $-80280.59$   \\  
$3p~^2P^o_{3/2}$   & $-92976.75$ & $-88181.44$ & $-83912.52$ & $-83609.01$ & $-81160.91$ & $-81161.53$ & $-80317.55$ & $-80280.59$    \\ 
$4s~^2S_{1/2}$     & $-92523.28$ & $-87803.36$ & $-83535.49$ & $-83233.24$ & $-80781.16$ & $-80783.73$ & $-79938.22$ & $-79904.88$  \\
$3d~^2D_{3/2}$     & $-92692.29$ & $-87950.83$ & $-83681.82$ & $-83371.89$ & $-80917.48$ & $-80913.77$ & $-80068.53$ & $-80046.22$    \\
$3d~^2D_{5/2}$     & $-92693.54$ & $-87951.34$ & $-83682.35$ & $-83372.58$ & $-80918.17$ & $-80914.49$ & $-80069.25$ & $-80046.94$      \\ 
$4p~^2P^o_{1/2}$   & $-92719.55$ & $-87937.46$ & $-83671.22$ & $-83367.50$ & $-80915.23$ & $-80916.80$ & $-80071.81$ & $-80043.38$    \\
$4p~^2P^o_{3/2}$   & $-92718.58$ & $-87937.53$ & $-83671.32$ & $-83367.58$ & $-80915.31$ & $-80916.86$ & $-80071.89$ & $-80043.46$   \\ 
\hline \hline
\end{tabular}
\label{tab_SMSGRASP}
\end{table*}

\subsection{Approaches to estimate the IS constants} 

All the considered three IS operators are scalar and correspond to the first-order energy corrections. Thus, they can be estimated by adopting the FF approach or by calculating expectation values of the respective IS operators \cite{bijaya-review, bijaya-li}. As shown in Eq. (\ref{propeq}), the expectation value evaluation (EVE) procedure in our RCC method contains two non-terminating series. As a result, this approach is not suitable to estimate the IS constants accurately as these quantities are more sensitive to higher-order correlations, particularly the SMS constants, comparative to the calculations of the E1 matrix elements and $A_{hf}$ values. Again, the FF approach is straightforward to adopt in any many-body method that gives energies. Thus, we consider this approach for both the MCDHF/RCI and RCC methods to estimate the IS constants. 

In the FF approach, an effective Hamiltonian is defined as $H_o=H +\lambda_o O$ to estimate the IS factors with $O=\sum_i o(r_i)$ denoting either the FS or MS operators with $\lambda_o$ as an arbitrary parameter satisfying the dimension correctly. For brevity, calculations are carried out in a.u. and the results are converted to other units as per the requirement. Thus, an IS factor is extracted in the FF approach as \cite{bijaya-review}
\begin{eqnarray}
\langle O \rangle \simeq \frac{E_v(+\lambda_o) - E_v(-\lambda_o)}{2 \lambda_o} ,
\label{eqff}
\end{eqnarray}
where $E_v(\pm \lambda_o)$ are the calculated total energies at the $\pm \lambda$ values using the Hamiltonian $H_o$.  In principle, the optimal choice of $\lambda_o$ can be state-dependent as the magnitude of $O$ can vary by one or two orders for different states. However, for practical reasons, we have considered the same value $\lambda_o=10^{-5}$~a.u. for all calculated states. 

\begin{table*}[t!]
\centering
\caption{Calculated IS constants at different levels of the approximation in the RCC method. The NMS constants are compared with the scaled values (Scaling) that are obtained from the ionization potentials listed in the NIST ADS \cite{NISTdata}.}
\begin{tabular}{c ccc  ccc  cccc } \hline \hline
&  \multicolumn{3}{c}{Finite-field approach} & \multicolumn{3}{c}{EVE approach} & \multicolumn{3}{c}{Analytical response approach} \\
State  & DHF   &   RCCSD  & RCCSDT  & DHF & RCCSD & RCCSDT & DHF  & RCCSD  & RCCSDT  & Scaling \cite{NISTdata} \\ \hline
  & & & & & & & & & & \\
\multicolumn{11}{c}{FS constants MHz fm$^{-2}$ } \\
$3s~^2S_{1/2}$     &   $-106.96$ & $-116.71$ &  $-117.09$  & $-104.57$  & $-116.12$ & $-116.75$ & $-104.57$  & $-116.23$ & $-116.85$  & \\
$3p~^2P^o_{1/2}$   &   9.15 & 9.66  &  9.62  & $-0.06$ & 10.02 &  10.01 & $-0.06$ &   9.87  & 9.80  & \\
$3p~^2P^o_{3/2}$   &   9.57 & 9.98  &  9.93  & $\sim 0.0$ &  10.02 & 9.98 & $\sim 0.0$ & 9.88   & 9.82  & \\
$4s~^2S_{1/2}$     &   $-29.92$  & $-31.86$  & $-31.91$ & $-29.45$ & $-31.70$ &  $-31.80$ & $-29.45$   & $-31.69$   & $-31.80$ & \\
$3d~^2D_{3/2}$     &   $-0.06$   & $-0.19$   & $-0.19$   & $\sim 0.0$ & 0.02 & $-0.09$ & $\sim 0.0$ &  $-0.06$ & $-0.14$ & \\
$3d~^2D_{5/2}$     &   $-0.16$   & $-0.33$   & $-0.33$   & $\sim 0.0$ & 0.02 & $-0.09$ & $\sim 0.0$ &  $-0.06$ & $-0.14$ & \\
$4p~^2P^o_{1/2}$   &   3.15      & 3.30  &  3.29  & $-0.02$ & 3.38 & 3.39 & $-0.02$    &  3.34   & 3.32 & \\
$4p~^2P^o_{3/2}$   &   3.34      & 3.45 &  3.43  & $\sim 0.0$ & 3.38 & 3.38 & $\sim 0.0$ & 3.34  & 3.33 \\
  & & & & & & & & & & \\
\multicolumn{11}{c}{NMS constants in GHz u} \\
$3s~^2S_{1/2}$     &  1952.40  & 1991.25 &  1992.35  & 2649.87 & 2013.05 & 1976.63 & 2649.87  & 1990.28  & 1978.31  & 1994.37 \\
$3p~^2P^o_{1/2}$   &  1385.33  & 1405.97 &  1406.61  & 1854.16 & 1419.88 & 1398.21 & 1854.16  & 1406.36 & 1400.42   & 1407.75 \\
$3p~^2P^o_{3/2}$   &  1384.16  & 1404.77 & 1405.42 & 1851.86 & 1408.89 & 1389.46 & 1851.86  & 1405.52  & 1399.88  &  1406.25  \\
$4s~^2S_{1/2}$     &  835.91   & 845.60  &  845.84  & 1025.27 & 848.10 & 837.00 & 1025.27  & 845.82  &  842.45  & 846.36 \\
$3d~^2D_{3/2}$     &   811.47  & 818.39 &  818.54  & 857.58 & 812.09 & 809.20 & 857.58 & 815.46 & 815.75 &  818.64  \\
$3d~^2D_{5/2}$     &   811.34  & 818.18 &  818.34  & 857.81 & 812.39 & 811.98 & 857.81   & 815.58  & 815.89  &  818.63 \\
$4p~^2P^o_{1/2}$   &   661.55  & 667.88 &  668.07  & 821.36 & 670.12 & 660.83 & 821.36   & 668.36  & 666.19   &  668.50 \\
$4p~^2P^o_{3/2}$   &   661.24  & 667.58  & 667.77  & 820.65 & 669.82 & 661.67 &  820.65  & 668.19 & 665.98  &  668.00 \\
  & & & & & & & & & & \\
\multicolumn{11}{c}{SMS constants in GHz u} \\
$3s~^2S_{1/2}$     &  $-179.60$  & 111.34 &  46.03  & $-564.23$ & 47.04 & 65.05 & $-564.23$ & 111.89  & 50.18  & \\
$3p~^2P^o_{1/2}$   & $-396.16$  & $-276.20$  &  $-310.19$ & $-592.61$ & $-319.66$ & $-314.98$ & $-592.61$  & $-276.36$  & $-308.65$ & \\
$3p~^2P^o_{3/2}$   &  $-400.05$  & $-280.93$   & $-314.74$  & $-594.96$  & $-324.81$ & $-319.91$ & $-594.96$  & $-280.76$   & $-312.71$ & \\
$4s~^2S_{1/2}$     &  $-33.13$   & 55.63 &  42.92  & $-137.09$ & 38.22 & 43.65 & $-137.09$  & 55.61   &  41.75 & \\
$3d~^2D_{3/2}$     &  $-100.29$  & $-99.37$  & $-109.81$ & $-119.97$ & $-98.42$ & $-98.29$ & $-119.97$  & $-98.07$ & $-109.05$ & \\
$3d~^2D_{5/2}$     &  $-100.39$  & $-99.56$  & $-110.00$ & $-120.73$& $-99.16$ & $-99.10$ & $-120.73$  & $-98.77$  & $-109.75$ & \\
$4p~^2P^o_{1/2}$   & $-137.85$  &  $-94.50$  &  $-104.42$ & $-204.77$ & $-110.02$ & $-107.99$ & $-204.77$  & $-94.83$  & $-104.10$ & \\
$4p~^2P^o_{3/2}$   &  $-139.31$  & $-96.22$ &  $-106.11$ & $-205.68$ & $-111.83$ & $-109.69$ & $-205.68$   &  $-96.39$ &  $-105.57$ & \\
\hline \hline
\end{tabular}
\label{tab_sms}
\end{table*}

{\it Albeit} the FF approach is a universally accepted approach to determine the IS constants, it is possible that the results for some states may suffer numerical errors due to arbitrariness in the choice of $\lambda_o$ and non-negligible contributions from the non-linear IS contributions to total energies $E_v(\pm \lambda_o)$. In order to avoid this problem and to validate the results, we also use both the EVE and AR approaches in the RCC theory to evaluate the IS constants that do not depend on any arbitrary parameter. As discussed before, the EVE contains two non-terminating series. We give results from the EVE procedure by terminating the non-terminating series of the RCC expressions forcefully at the same level of the calculations of the E1 matrix elements and $A_{hf}$ values for completeness. However, by applying brute-force truncation of these series make it challenging to estimate the SMS constants accurately, which are evaluated using a two-body operator. In contrast to the FF, both the EVE and AR approaches do not include the orbital relaxation effects implicitly. The AR approach contains all terminating expressions, the orbital relaxation effects are accounted through the perturbation in this approach. Thus, some of these effects can be missed in a truncated RCC theory at the same level of approximation with the FF approach. 

In the AR approach, an atomic wave function is expanded as \cite{bijaya-review, bijaya-ar}
\begin{eqnarray}
|\Psi_v \rangle &=&  |\Psi_v^{(0)} \rangle + \lambda_o |\Psi_v^{(1)} \rangle  + \cdots .
\end{eqnarray}
The first-order wave function solving equation is given by
\begin{eqnarray} 
 \left ( H_\text{at} - E_v^{(0)}\right ) |\Psi_v^{(1)} \rangle = \left ( \langle O \rangle - O \right ) |\Psi_v^{(0)} \rangle .
 \label{eqar}
\end{eqnarray}

In the RCC theory framework, the zeroth and first-order perturbed wave functions for the closed-core can be defined as
\begin{eqnarray}
|\Psi_0^{(0)} \rangle = e^{T^{(0)}} |\Phi_0 \rangle  
\end{eqnarray}
and
\begin{eqnarray}
|\Psi_0^{(1)} \rangle = e^{T^{(0)}} T^{(1)} |\Phi_0 \rangle ,
\end{eqnarray}
where $T^{(0)}$ and $T^{(1)}$ are the zeroth- and first-order RCC operators respectively. Similarly, these wave functions including the valence orbital can be written as
\begin{eqnarray}
|\Psi_v^{(0)} \rangle = e^{T^{(0)}} \left (1+S_v^{(0)}) \right )  | \Phi_v \rangle 
\end{eqnarray}
and 
\begin{eqnarray}
|\Psi_v^{(1)} \rangle = e^{T^{(0)}} \left ( S_v^{(1)} + (1+S_v^{(0)})T^{(1)} \right )  | \Phi_v \rangle , 
\end{eqnarray}
where $S_v^{(0)}$ and $S_v^{(1)}$ are the zeroth- and first-order open-shell RCC operators respectively. 

The amplitude-determining equations for the first-order RCC operators are given by \cite{bijaya-ar}
\begin{eqnarray}
&& \langle \Phi_0^* |  \left ( H_\text{at} e^T T^{(1)} + O e^{T^{(0)}} \right )_l  | \Phi_0 \rangle = 0 \label{eqs01}\\
& \text{and} & \nonumber \\
&&  \langle \Phi_v^* | \left \{ \left ( H_\text{at} e^{T^{(0)}} \right )_l -E_v^{(0)})  \right \} S_v^{(1)}  + \left ( H_\text{at} e^{T^{(0)}} T^{(1)} \right )_l \nonumber \\
&&   \times  \left \{ 1+ S_v^{(0)} \right \} + \left ( O e^{T^{(0)}} \right )_l \left \{ 1+ S_v^{(0)} \right \} \nonumber \\
&& + \langle O \rangle S_v^{(0)} | \Phi_v \rangle = 0 .  \ \ \ \label{eqv1} \ \ 
\label{eqamp1}
\end{eqnarray}

This follows the IS is evaluating expression as 
\begin{eqnarray}
\langle O \rangle  &=& \langle \Phi_v | \left ( H_\text{at} e^{T^{(0)}} \right )_l S_v^{(1)}  + \left ( H_\text{at} e^{T^{(0)}} T^{(1)} \right )_l  \left \{ 1+ S_v^{(0)} \right \}   \nonumber \\
&& + ( O e^{T^{(0)}})_l \left \{1 + S_v^{(0)} \right \} | \Phi_v \rangle . 
\label{eqeng1}
\end{eqnarray}

\begin{table*}[t!]
    \centering
    \caption{The differential IS constants of the low-lying transitions in Mg$^+$ from the MCDHF/RCI method.}
    \begin{tabular}{c |cc |cc |cc } \hline \hline
       &  \multicolumn{2}{|c}{$F$ values (in MHz/fm$^2$)} & \multicolumn{2}{|c}{$K^{NMS}$ values (in GHz amu)} & \multicolumn{2}{|c}{$K^{SMS}$ values (in GHz amu)} \\
    \cline{2-7}  \vspace{-0.25cm}
    &&&&&&\\
       Transition                                 & SD       & SDT      & SD  & SDT  & SD & SDT \\ \hline
       $3s~^2S_{1/2} \rightarrow 3p~^2P^o_{1/2}$  &  $-125$  & $-125$   &  613 &  596 & 304   & 321   \\
       $3s~^2S_{1/2} \rightarrow 3p~^2P^o_{3/2}$  &  $-125$  &  $-125$  &  614 &  597 & 304   & 321   \\
       $3s~^2S_{1/2} \rightarrow 4s~^2S_{1/2}$    &  $-84.5$ &  $-84.1$ & 1205 & 1162 & $-75$ & $-54$ \\ 
       $3s~^2S_{1/2} \rightarrow 3d~^2D_{3/2}$    &  $-116$  &  $-116$  & 1249 & 1196 & 55    & 87    \\
       $3s~^2S_{1/2} \rightarrow 3d~^2D_{5/2}$    &  $-116$  &  $-116$  & 1249 & 1196 & 55    & 88    \\
       $3s~^2S_{1/2} \rightarrow 4p~^2P^o_{1/2}$  &  $-119$  &  $-119$  & 1387 & 1342 & 58    & 84    \\
       $3s~^2S_{1/2} \rightarrow 4p~^2P^o_{3/2}$  &  $-119$  &  $-119$  & 1387 & 1342 & 58    & 84    \\
       \hline \hline
    \end{tabular}
    \label{tab_ISGRASP}
\end{table*}

\section{Results and Discussions}

As a starting point, an MCDHF calculation was performed in the extended optimal level scheme for the weighted average of the targeted states of the $3s,4s,3d$ even and $3p,4p$ odd parity configurations. In this initial calculation, to obtain additional spatially extended orbitals as required to ensure a good agreement between transition rates in the Coulomb and Babaushkin gauges \cite{atoms7040106,Pehlivan}, states belonging to the $5s,6s,7s,4d$ even and $5p,6p,7p$ odd parity configurations were also added. Keeping the orbitals of the above states fixed, six layers of correlation orbitals were determined from subsequent MCDHF calculations based on expansions accounting for valence and core-valence correlation effects. The latter was obtained by allowing single- and double (SD) substitutions from the  $3s,4s,3d$ even and $3p,4p$ odd parity configurations with the restriction that there is at most one substitution from the $1s^22s^22p^6$ atomic core. To include core-core correlating orbitals in the orbital basis a seventh orbital layer was determined based on an expansion for which all SD substitutions were allowed. The orbitals of this layer are considerably more contracted than the ones of the first six layers determined based on expansions accounting for valence and core-valence correlation effects.

The seven layers of correlation orbitals are shown in Table \ref{tab:layers} in non-relativistic notation. The above computational approach is static in the sense that the orbitals of the targeted states are frozen. To account for the modification of the orbitals of the targeted states by correlation corrections, the orbitals were transformed to natural orbitals, see \cite{PhysRevA.101.062510}, based on a calculation for which the CSF expansion was obtained by allowing only S substitutions. The natural orbitals were used in the subsequent RCI calculations. For the RCI calculations, the CSF expansions were obtained from SD substitutions from the reference configurations to the increasing layers of orbitals and thus they also account for core-core correlation. Finally, to include three-particle effects, the SD multi-reference expansion based on layer 7 was augmented with CSFs obtained by triple (T) substitutions from the reference configurations to the $\{12s,10p,5d\}$ orbital set. The MCDHF and RCI calculations were performed with {\sc Graspg} \cite{SI2025109604}, an extension of Grasp2018 \cite{FroeseFischer.2019.p184} based on CSFs. Transformation to natural orbitals was done with the {\tt RDENSITY} program \cite{SCHIFFMANN2022108403}. The isotope shift parameters were evaluated using {\tt RIS4} \cite{EKMAN2019433}.

In the RCC calculations, the DHF orbitals were constructed using a large number of Gaussian-type orbitals that satisfy the even tempering condition. The kinetic balance condition was also ensured between the large and small components while obtaining these orbitals. We considered $19s$, $19p$, $19d$, $18f$, $16g$, $13h$, and $11i$ orbitals in the RCCSD method to carry out the calculations. However, including all these orbitals in the RCCSDT method approximation was a challenge so we only included orbitals up to $g-$symmetry orbitals in this method. To demonstrate the importance of the contributions from the triples excitations, we first performed calculations by considering orbitals up to $g-$ symmetry in the RCCSD and RCCSDT methods. For the final results, the extra RCCSD contributions from the $h-$ and $i-$ symmetry orbitals are given as corrections under `Basis'. We also estimated contributions from the Breit interactions (given as `Breit') and QED effects (given as `QED') approximation, allowing orbitals up to $g-$symmetry in the RCCSD method. 

\begin{table*}[t!]
    \centering
    \caption{Corrections to the IS constants at the RCCSD method approximation.}
    \begin{tabular}{c |ccc |ccc |ccc } \hline \hline
       &  \multicolumn{3}{|c}{$F$ values (in MHz/fm$^2$)} & \multicolumn{3}{|c}{$K^{NMS}$ values (in GHz amu)} & \multicolumn{3}{|c}{$K^{SMS}$ values (in GHz amu)} \\
    \cline{2-7}  \vspace{-0.25cm}
    &&&&&&\\
 State & Basis  &   Breit  & QED  & Basis  &   Breit  & QED  & Basis  &   Breit  & QED   \\ \hline
       $3s~^2S_{1/2}$  & $-0.48$  &  0.19  &  0.51 & $\sim 0.0$ & $-0.32$  & $-0.57$ & $\sim 0.0$  & 0.97 & $0.03$ \\
       $3p~^2P^o_{1/2}$  & $\sim 0.0$ & $-0.04$ &  $0.46$ & $\sim 0.0$ & $-0.57$ & $-0.28$ & $\sim 0.0$ &  0.69 & $0.16$\\
       $3p~^2P^o_{3/2}$  & 0.01 & $-0.45$  & $-0.26$ & $\sim 0.0$ & $0.16$ & $0.13$ & $\sim 0.0$ & 0.51 & $0.14$ \\
       $4s~^2S_{1/2}$  & $-0.17$ & 0.13 & 0.30 & $\sim 0.0$ & $-0.17$  & $-0.27$ & $\sim 0.0$ & 0.30 & 0.03 \\ 
       $3d~^2D_{3/2}$  & $\sim 0.0$  & $0.24$ & $0.35$ & 0.08 & $0.18$ & $-0.06$ & $\sim 0.0$ & $0.14$ & 0.41 \\
       $3d~^2D_{5/2}$  & $\sim 0.0$  & $0.14$ & $0.12$ & 0.08 & 0.15 & $0.27$ & $\sim 0.0$ & $-0.25$ & $-0.29$ \\
       $4p~^2P^o_{1/2}$  & $\sim 0.0$ & $\sim 0.0$ & $0.11$ & $\sim 0.0$ & $-0.15$ & $-0.04$ & $\sim 0.0$ & 0.08 & $-0.12$\\
       $4p~^2P^o_{3/2}$  & $\sim 0.0$ & $-0.06$ & $-0.12$ & $\sim 0.0$ & $-0.11$ & $-0.26$ & $\sim 0.0$  & 0.17  & $-0.01$\\
       \hline \hline
    \end{tabular}
    \label{tab_ISCor}
\end{table*}

Several theoretical investigations of properties calculated in this work have already been carried out using the MCDHF/RCI and RCC methods or their variants. Since our interest is to make a comparative analysis of the results of different properties from both the MCDHF/RCI and RCC methods, our calculations are compared only with the precisely available experimental values. We discuss the below excitation energies, E1 matrix elements, $A_{hf}$, and IS constants of the low-lying states of Mg$^+$ from both methods. First, we give results from the MCDHF/RCI method and they are then compared with the RCC calculations.

\subsection{Energies}

The excitation energies from the MCDHF/RCI method are displayed in Table \ref{tab_exeCI} as functions of the layers of correlation orbitals. 
The last row reports the number of CSFS, NCSFs, for each layer. As can be seen from the table, the change in excitation energies overall is small, and after 4 layers the energies are stable within 1 cm$^{-1}$. The excitation energies are all too low, indicating that there is relatively more uncaptured correlation energy in the ground state relative to the excited states. To quantify this, we evaluate the mean level deviation (MLD) 
\begin{equation}
\text{MLD} = \frac{1}{N}\sum_{i=1}^{N}|E_{obs}(i)-E_{calc}(i)+ES|,
\end{equation}
where values from the NIST ADS \cite{NISTdata} are used as $E_{obs}$. The energy shift (ES) is chosen as to minimize the sum and reflects the degree to which the ground state level is either favored or disfavored in the theoretical balance of binding energy. The obtained MLD is $22$ $\rm{cm^{-1}}$  based on  $ES = -183$ $\rm{cm^{-1}}$. Including triple excitations increases the MLD $49$ $\rm{cm^{-1}}$  but lowers the energy shift to $ES = -149$ $\rm{cm^{-1}}$. Further improvements of the energies would result from the inclusion of orbitals with higher angular momenta. Also, the inclusion of the neglected three- and four-particle effects would bring the excitation energies in better agreement with experiment.  

The second ionization potential of the ground state and excitation energies of the other states of Mg$^+$ from the RCC method are presented in Table \ref{tab_execc}. As can be seen, the energy values are gradually increasing from the DHF to RCCSD to RCCSDT methods. The corrections due to `Basis', `Breit', and `QED' are found to be substantially small and almost get canceled out with each other. The final values are taken as the RCCSDT values with other corrections and their uncertainties are quoted by analyzing the differences in the RCCSDT and RCCSD values and other neglected effects. Comparison between the RCC and experimental values \cite{NISTdata} shows good agreement. By comparing results from \mbox{layer 7} and \mbox{layer 7$+T$} with the RCCSD and RCCSDT calculations, we find that the RCC values are closer to the experimental results. However, it should be noted that the RCC theory at the RCCSD approximation also contains non-linear terms corresponding to triple and higher-level excitations. Therefore, we hope that after considering full triple excitations, the MCDHF/RCI results can improve further.

\subsection{E1 amplitudes}

The transition rates in length and velocity forms are displayed in Table \ref{tab_A} as functions of the layers of correlation orbitals. The agreement between the rates in the two gauges is good, with the average relative difference being less than 0.5 \%. Taking the full behavior of the gauge parameter $G$ and analyzing the values according to the quantitative and qualitative approach \cite{Rynkun.2022.p82,Kitoviene.2024.p}, all the transitions were found to be in the NIST ASD AA class, uncertainty less than $1\%$.  The transition matrix elements are shown in Table \ref{tab_matE1} in the length form. Here we note that the matrix for $3s~^2S_{1/2} \rightarrow 4p~^2P^o_{1/2}$ is two orders of magnitude smaller than the other matrix elements. As explained in \cite{MCHF} Sec. 9.7, this is due to extensive radial cancellation in the dominating transition integral. The transition integrand has a negative and a positive part, which almost perfectly cancels yielding an integral close to zero. This extensive radial cancellation makes this transition amplitude sensitive to correlation effects as can be seen from the relatively large change when going from layer 1 to layer 7. From the transition rates, the lifetimes of the states and the branching fractions can be inferred. 

The E1 matrix elements using the length gauge expression at different levels of approximation in the RCC method are listed in Table \ref{tab_E1}. When comparing these values with the E1 matrix elements from Table \ref{tab_matE1} of the MCDHF/RCI method, we find differences in the second decimal places from both methods. In order to find out the significance of these differences, we use them to estimate the lifetimes of the excited states and $\alpha_d$ value of the ground state by combining then with the experimental energies. The estimated lifetimes of the excited states using E1 matrix elements from the MCDHF/RCI and RCC methods are given in Table \ref{tab_LMCDHF} along with the BR values. Uncertainties to the RCC values are also estimated from the uncertainties of the E1 matrix elements. We find reasonably good agreement between the estimated lifetimes, $\tau$, using matrix elements from both methods. We find that the available experimental $\tau$ values of the considered excited states in Mg$^+$ are less precise and the central values of the measurements differ a lot \cite{Ansbacher1989, Lundin1973, Schaefer1971, Andersen1970, Berry1970}. Consistent in the $\tau$ values for all the investigated states from both methods show that their estimated values are very reliable and more precise than the available experimental results. Therefore, we believe that our recommended lifetime values can be useful for their measurements as well as for other applications.

In addition to testing the reliability of the calculated E1 matrix elements from both methods by estimating the lifetimes of atomic states, we intend to use them to determine the $\alpha_d$ value of the ground state. The $\alpha_d$ values obtained using the E1 matrix elements from the MCDHF/RCI and RCC methods in a sum-over-states approach are given in Table \ref{tab_pol}. To get the final values, we have used contributions from the core correlation, core-valence correlation, and high-lying states (given as ``Tail") from random phase approximation (RPA). We find both the results from the MCDHF/RCI and RCC methods agree very well, they differ slightly from the sole experimental value available in the literature \cite{Mgionpol}.  

\begin{table*}[t!]
    \centering
    \caption{Recommended IS constants of low-lying transitions in Mg$^+$ from the RCC theory. They are considered as the sum of RCCSDT values with the corrections listed in the previous table and compared with the experimental values extracted from the King's plot.}
    \begin{tabular}{c |cc |cc |cc } \hline \hline
       &  \multicolumn{2}{|c}{$F$ values (in MHz/fm$^2$)} & \multicolumn{2}{|c}{$K^{NMS}$ values (in GHz amu)} & \multicolumn{2}{|c}{$K^{SMS}$ values (in GHz amu)} \\
    \cline{2-7}  \vspace{-0.25cm}
    &&&&&&\\
       Transition  & This work  & Experiment  & This work  & Scaling  & This work & Experiment \\ \hline
       $3s~^2S_{1/2} \rightarrow 3p~^2P^o_{1/2}$  & $-126(1)$  & $-127(12)$ \cite{2012-Yor} & 586(2)  &  586.59 & 356(15) & 369.3(3) \cite{2009-MgTrap} \\
       $3s~^2S_{1/2} \rightarrow 3p~^2P^o_{3/2}$  & $-126(1)$  &  & 587(2) &  588.09 & 362(15) & 367.7(3) \cite{2009-MgTrap} \\
       $3s~^2S_{1/2} \rightarrow 4s~^2S_{1/2}$  &  $-85(1)$  &  & 1147(1) & 1147.98 & 3(5) & \\ 
       $3s~^2S_{1/2} \rightarrow 3d~^2D_{3/2}$  &  $-116(1)$ &  & 1174(1) & 1175.70 & 156(10) &  \\
       $3s~^2S_{1/2} \rightarrow 3d~^2D_{5/2}$  &  $-116(1)$ &  & 1174(1) & 1175.71 & 156(10) & \\
       $3s~^2S_{1/2} \rightarrow 4p~^2P^o_{1/2}$  & $-120(1)$  &  & 1324(1) & 1325.84 & 151(10) & \\
       $3s~^2S_{1/2} \rightarrow 4p~^2P^o_{3/2}$  &  $-120(1)$ &  & 1325(1) & 1326.34 & 153(10) & \\
       \hline \hline
    \end{tabular}
    \label{tab_IS}
\end{table*}

\subsection{$A_{hf}$ values}

The $A_{hf}$ values from the MCDHF/RCI method are displayed in Table \ref{tab_hfs_MCDHF} as functions of the layers of correlation orbitals. Hyperfine structure constants are known to be relatively sensitive to correlation effects, and this is seen in the table, where the values change markedly when going from layer 1 to layer 7. We now also see that the triple substitutions are important for the hyperfine constants at the current level of accuracy. The hyperfine constant for the $3s~^2S_{1/2}$ ground state was recently calculated using the MCDHF/RCI method yielding the result $-595.262$ MHz in better agreement with experiment \cite{Xu2017,Itano1981}. Both sets of calculations were based on natural orbitals, including the same correlation effects. The difference in result can be attributed to the fact that the current orbital set, spanning many states, is less complete.

The RCC values for $A_{hf}$ at different approximations are given in Table \ref{tab_hfs}. We have also accounted for the Bohr-Weisskopf (BW) effects to the estimation of these quantities using the RCCSD method. The table shows that the estimated $+$Basis, $+$Breit, and $+$QED corrections are quite small except in the ground state. The QED correction helps to match the calculated RCC value with the measurements \cite{Xu2017,Itano1981} even without using the natural orbitals. Comparing MCDHF/RCI and RCC results, we find that they agree reasonably, but the RCC values are expected to be more precise on the basis of the ground-state results.

\subsection{Isotope shift constants}

Having learned about the correlation trends in the energies, E1 matrix elements, and $A_{hf}$ values, we now turn to discuss the IS constants. Correlation trends in the FS, NMS and SMS constants are expected to behave differently. The FS constants from the MCDHF/RCI method are displayed in Table \ref{tab_SMSGRASP} as functions of the layers of correlation orbitals. The change in the values between layer 5 and layer 6 is very small. However, for layer 7 the densities increase mainly due to the inclusion of the contracted orbitals optimized on core-core effects. The effects of the triple substitutions are small. The convergence of the NMS and SMS parameters are also displayed in Table \ref{tab_SMSGRASP}. Again, there is an appreciable change in the parameters when the contracted orbitals from layer 7 are included. It would be desirable to add more orbital layers optimized on core-core correlation to establish convergence. However, the current limitation of the number of allowed orbitals in {\sc Grasp} prevents this. Since the differential IS constants of the atomic transitions are of actual experimental interest, we present these values from SD and SDT approximations of the MCDHF/RCI method in Table \ref{tab_ISGRASP} for the low-lying transitions of Mg$^+$. This table shows that triple excitation contributions are negligible in the determination of the FS parameters while they are quite significant in the evaluation of the NMS and SMS constants. 
For the SMS constants, being due to a two-body operator, the effects of quadruple substitutions are expected to be non-negligible.

The IS constants obtained at different levels of approximation in the RCC method are presented in Table \ref{tab_sms}. These values for each state and parameter look completely different than the MCDHF/RCI results. This is owing to the fact that we have not included core contributions in the RCC calculations, which are common to all the considered states and can cancel out while determining the differential value of a transition. In order to understand the roles of electron correlation effects and the dependency of the results on the approach adopted to estimate the IS parameters through the RCC theory, we give results from the FF, EVE, and AR approaches at the DHF, RCCSD, and RCCSDT approximations in the above table. As can be seen, the FS constants are almost consistent in all three approaches but the NMS and SMS constants mostly differ. The NMS constants from the RCCSDT method of the FF approach match better with the scaled values from the experimental energies \cite{NISTdata} (given in the table as ``scaling") than the other two approaches. Based on this finding, we presume that the RCCSDT values of the SMS constants from the FF approach are more accurate. Also, the reasonable agreement among the SMS constants from both the FF and AR approaches supports further our assessment.  

To give the final values of the IS constants from the RCC method, we also estimate the Basis, Breit, and QED corrections to these constants at the RCCSD method using the FF approach and these corrections are listed in Table \ref{tab_ISCor}. By considering these corrections with the RCCSDT values of the IS constants from the FF approach, we give the final differential IS values of the low-lying transitions from the RCC theory in Table \ref{tab_IS} along with their uncertainties. Comparing results from both Tables \ref{tab_ISGRASP} and \ref{tab_IS}, we find that the FS constants from both the MCDHF/RCI and RCC methods match well and also the value for the D1 line is well within the quoted uncertainty of the lonely available experimental value \cite{2012-Yor}. The differential NMS constants from the RCC method are more close to the scaling values than the MCDHF/RCI values. There are, however, large differences among the differential SMS constants from the MCDHF/RCI and RCC methods seen. Particularly, the sign of the central value of SMS constant of the $3s~^2S_{1/2} \rightarrow 4s~^2S_{1/2}$ transition is opposite from both methods. The experimental values of the SMS constants of the D1 and D2 lines \cite{2009-MgTrap} are closer to the RCC results than the MCDHF/RCI method. We anticipate that the inclusion of triple excitations in the MCDHF/RCI method can bring the NMS and SMS constants closer to the experimental values. In fact, the quadruple excitation contributions can further improve the RCC results.  

\section{Summary}

We have employed the MCDHF/RCI and RCC methods to calculate energies, electric dipole matrix elements, magnetic dipole hyperfine structure constants and isotope shift constants of low-lying states of Mg$^+$. The MCDHF/RCI calculations are carried out at different layers by increasing the size of active space gradually in order to understand the roles of the high-lying orbitals in the determination of properties of Mg$^+$. In particular, we have identified the need to remove the limitations of the number of allowed radial orbitals in the current codes to add more orbitals optimized on core-core correlation.

The RCC calculations are carried out at the singles and doubles excitation approximation as well as at the singles, doubles, and triples excitation approximation. The differences in the results from both the approximations demonstrate the importance of the triple excitations in Mg$^+$. Comparisons between the MCDHF/RCI and RCC results are made for each property to understand their capabilities to produce accurate results. This exercise would be useful in understanding possible discrepancies among the results of various properties from both methods when applied to other atomic systems. Consistencies in the estimated lifetimes of the excited states of Mg$^+$ from both the MCDHF/RCI and RCC methods suggest that they are very reliable and more precise than the available experimental results. The estimated ground state electric dipole polarizability from both methods agree quite well but differ from the measurement. This suggests for further probe in the preciseness of the measured electric dipole polarizability of Mg$^+$. Even though the MCDHF/RCI and RCC results match well in Mg$^+$, it would be interesting to make such a comparison of results in the heavier atomic systems.

\section*{Acknowledgment}

BKS acknowledges ANRF grant no. CRG/2023/002558 and Department of Space, Government of India for financial supports. The RCC calculations were carried out using the ParamVikram-1000 HPC of the Physical Research Laboratory (PRL), Ahmedabad, Gujarat, India.   PJ. acknowledges support from the Swedish Research Council (VR 2023-05367).

\bibliography{references}

\begin{thebibliography}{72}%
\makeatletter
\providecommand \@ifxundefined [1]{%
 \@ifx{#1\undefined}
}%
\providecommand \@ifnum [1]{%
 \ifnum #1\expandafter \@firstoftwo
 \else \expandafter \@secondoftwo
 \fi
}%
\providecommand \@ifx [1]{%
 \ifx #1\expandafter \@firstoftwo
 \else \expandafter \@secondoftwo
 \fi
}%
\providecommand \natexlab [1]{#1}%
\providecommand \enquote  [1]{``#1''}%
\providecommand \bibnamefont  [1]{#1}%
\providecommand \bibfnamefont [1]{#1}%
\providecommand \citenamefont [1]{#1}%
\providecommand \href@noop [0]{\@secondoftwo}%
\providecommand \href [0]{\begingroup \@sanitize@url \@href}%
\providecommand \@href[1]{\@@startlink{#1}\@@href}%
\providecommand \@@href[1]{\endgroup#1\@@endlink}%
\providecommand \@sanitize@url [0]{\catcode `\\12\catcode `\$12\catcode `\&12\catcode `\#12\catcode `\^12\catcode `\_12\catcode `\%12\relax}%
\providecommand \@@startlink[1]{}%
\providecommand \@@endlink[0]{}%
\providecommand \url  [0]{\begingroup\@sanitize@url \@url }%
\providecommand \@url [1]{\endgroup\@href {#1}{\urlprefix }}%
\providecommand \urlprefix  [0]{URL }%
\providecommand \Eprint [0]{\href }%
\providecommand \doibase [0]{https://doi.org/}%
\providecommand \selectlanguage [0]{\@gobble}%
\providecommand \bibinfo  [0]{\@secondoftwo}%
\providecommand \bibfield  [0]{\@secondoftwo}%
\providecommand \translation [1]{[#1]}%
\providecommand \BibitemOpen [0]{}%
\providecommand \bibitemStop [0]{}%
\providecommand \bibitemNoStop [0]{.\EOS\space}%
\providecommand \EOS [0]{\spacefactor3000\relax}%
\providecommand \BibitemShut  [1]{\csname bibitem#1\endcsname}%
\let\auto@bib@innerbib\@empty
\bibitem [{\citenamefont {Sahoo}\ \emph {et~al.}(2021)\citenamefont {Sahoo}, \citenamefont {Das},\ and\ \citenamefont {Spiesberger}}]{bijaya-pnc}%
  \BibitemOpen
  \bibfield  {author} {\bibinfo {author} {\bibfnamefont {B.~K.}\ \bibnamefont {Sahoo}}, \bibinfo {author} {\bibfnamefont {B.~P.}\ \bibnamefont {Das}},\ and\ \bibinfo {author} {\bibfnamefont {H.}~\bibnamefont {Spiesberger}},\ }\bibfield  {title} {\bibinfo {title} {New physics constraints from atomic parity violation in $^{133}\mathrm{Cs}$},\ }\href {https://doi.org/10.1103/PhysRevD.103.L111303} {\bibfield  {journal} {\bibinfo  {journal} {Phys. Rev. D}\ }\textbf {\bibinfo {volume} {103}},\ \bibinfo {pages} {L111303} (\bibinfo {year} {2021})}\BibitemShut {NoStop}%
\bibitem [{\citenamefont {Sahoo}\ \emph {et~al.}(2023)\citenamefont {Sahoo}, \citenamefont {Yamanaka},\ and\ \citenamefont {Yanase}}]{bijaya-xe}%
  \BibitemOpen
  \bibfield  {author} {\bibinfo {author} {\bibfnamefont {B.~K.}\ \bibnamefont {Sahoo}}, \bibinfo {author} {\bibfnamefont {N.}~\bibnamefont {Yamanaka}},\ and\ \bibinfo {author} {\bibfnamefont {K.}~\bibnamefont {Yanase}},\ }\bibfield  {title} {\bibinfo {title} {Revisiting theoretical analysis of the electric dipole moment of $^{129}\mathrm{Xe}$},\ }\href {https://doi.org/10.1103/PhysRevA.108.042811} {\bibfield  {journal} {\bibinfo  {journal} {Phys. Rev. A}\ }\textbf {\bibinfo {volume} {108}},\ \bibinfo {pages} {042811} (\bibinfo {year} {2023})}\BibitemShut {NoStop}%
\bibitem [{\citenamefont {Nataraj}\ \emph {et~al.}(2008{\natexlab{a}})\citenamefont {Nataraj}, \citenamefont {Sahoo}, \citenamefont {Das},\ and\ \citenamefont {Mukherjee}}]{bijaya-cs}%
  \BibitemOpen
  \bibfield  {author} {\bibinfo {author} {\bibfnamefont {H.~S.}\ \bibnamefont {Nataraj}}, \bibinfo {author} {\bibfnamefont {B.~K.}\ \bibnamefont {Sahoo}}, \bibinfo {author} {\bibfnamefont {B.~P.}\ \bibnamefont {Das}},\ and\ \bibinfo {author} {\bibfnamefont {D.}~\bibnamefont {Mukherjee}},\ }\bibfield  {title} {\bibinfo {title} {Intrinsic electric dipole moments of paramagnetic atoms: Rubidium and cesium},\ }\href {https://doi.org/10.1103/PhysRevLett.101.033002} {\bibfield  {journal} {\bibinfo  {journal} {Phys. Rev. Lett.}\ }\textbf {\bibinfo {volume} {101}},\ \bibinfo {pages} {033002} (\bibinfo {year} {2008}{\natexlab{a}})}\BibitemShut {NoStop}%
\bibitem [{\citenamefont {Mukherjee}\ \emph {et~al.}(2009)\citenamefont {Mukherjee}, \citenamefont {Sahoo}, \citenamefont {Nataraj},\ and\ \citenamefont {Das}}]{bijaya-fr}%
  \BibitemOpen
  \bibfield  {author} {\bibinfo {author} {\bibfnamefont {D.}~\bibnamefont {Mukherjee}}, \bibinfo {author} {\bibfnamefont {B.~K.}\ \bibnamefont {Sahoo}}, \bibinfo {author} {\bibfnamefont {H.~S.}\ \bibnamefont {Nataraj}},\ and\ \bibinfo {author} {\bibfnamefont {B.~P.}\ \bibnamefont {Das}},\ }\bibfield  {title} {\bibinfo {title} {Relativistic coupled cluster (rcc) computation of the electric dipole moment enhancement factor of francium due to the violation of time reversal symmetry},\ }\href {https://doi.org/10.1021/jp904020s} {\bibfield  {journal} {\bibinfo  {journal} {J. Phys. Chem. A}\ }\textbf {\bibinfo {volume} {113}},\ \bibinfo {pages} {12549} (\bibinfo {year} {2009})}\BibitemShut {NoStop}%
\bibitem [{\citenamefont {Nandy}\ and\ \citenamefont {Sahoo}(2015)}]{bijaya-dillip}%
  \BibitemOpen
  \bibfield  {author} {\bibinfo {author} {\bibfnamefont {D.~K.}\ \bibnamefont {Nandy}}\ and\ \bibinfo {author} {\bibfnamefont {B.~K.}\ \bibnamefont {Sahoo}},\ }\bibfield  {title} {\bibinfo {title} {Relativistic calculations of radiative properties and fine structure constant varying sensitivity coefficients in the astrophysically relevant zn\,ii, si\,iv and ti\,iv ions},\ }\href {https://doi.org/10.1093/mnras/stu2707} {\bibfield  {journal} {\bibinfo  {journal} {MNRAS}\ }\textbf {\bibinfo {volume} {447}},\ \bibinfo {pages} {3812} (\bibinfo {year} {2015})}\BibitemShut {NoStop}%
\bibitem [{\citenamefont {Sahoo}(2019)}]{bijaya-ca}%
  \BibitemOpen
  \bibfield  {author} {\bibinfo {author} {\bibfnamefont {B.~K.}\ \bibnamefont {Sahoo}},\ }\bibfield  {title} {\bibinfo {title} {High-precision determination of lorentz-symmetry-violating parameters in ${\mathrm{ca}}^{+}$},\ }\href {https://doi.org/10.1103/PhysRevA.99.050501} {\bibfield  {journal} {\bibinfo  {journal} {Phys. Rev. A}\ }\textbf {\bibinfo {volume} {99}},\ \bibinfo {pages} {050501} (\bibinfo {year} {2019})}\BibitemShut {NoStop}%
\bibitem [{\citenamefont {Sahoo}\ \emph {et~al.}(2025)\citenamefont {Sahoo}, \citenamefont {Blundell}, \citenamefont {Oleynichenko}, \citenamefont {Garcia~Ruiz}, \citenamefont {Skripnikov},\ and\ \citenamefont {Ohayon}}]{bijaya-review}%
  \BibitemOpen
  \bibfield  {author} {\bibinfo {author} {\bibfnamefont {B.~K.}\ \bibnamefont {Sahoo}}, \bibinfo {author} {\bibfnamefont {S.}~\bibnamefont {Blundell}}, \bibinfo {author} {\bibfnamefont {A.~V.}\ \bibnamefont {Oleynichenko}}, \bibinfo {author} {\bibfnamefont {R.~F.}\ \bibnamefont {Garcia~Ruiz}}, \bibinfo {author} {\bibfnamefont {L.~V.}\ \bibnamefont {Skripnikov}},\ and\ \bibinfo {author} {\bibfnamefont {B.}~\bibnamefont {Ohayon}},\ }\bibfield  {title} {\bibinfo {title} {Recent advancements in atomic many-body methods for high-precision studies of isotope shifts},\ }\href {https://doi.org/10.1088/1361-6455/adacc1} {\bibfield  {journal} {\bibinfo  {journal} {J. Phys. B: Atomic, Molecular and Optical Physics}\ }\textbf {\bibinfo {volume} {58}},\ \bibinfo {pages} {042001} (\bibinfo {year} {2025})}\BibitemShut {NoStop}%
\bibitem [{\citenamefont {Biero\ifmmode~\acute{n}\else \'{n}\fi{}}\ and\ \citenamefont {Pyykk\"o}(2001)}]{PhysRevLett.87.133003}%
  \BibitemOpen
  \bibfield  {author} {\bibinfo {author} {\bibfnamefont {J.}~\bibnamefont {Biero\ifmmode~\acute{n}\else \'{n}\fi{}}}\ and\ \bibinfo {author} {\bibfnamefont {P.}~\bibnamefont {Pyykk\"o}},\ }\bibfield  {title} {\bibinfo {title} {Nuclear quadrupole moments of bismuth},\ }\href {https://doi.org/10.1103/PhysRevLett.87.133003} {\bibfield  {journal} {\bibinfo  {journal} {Phys. Rev. Lett.}\ }\textbf {\bibinfo {volume} {87}},\ \bibinfo {pages} {133003} (\bibinfo {year} {2001})}\BibitemShut {NoStop}%
\bibitem [{\citenamefont {Sahoo}(2006)}]{bijaya-sr}%
  \BibitemOpen
  \bibfield  {author} {\bibinfo {author} {\bibfnamefont {B.~K.}\ \bibnamefont {Sahoo}},\ }\bibfield  {title} {\bibinfo {title} {Determination of the nuclear quadrupole moment of $^{87}\mathrm{Sr}$},\ }\href {https://doi.org/10.1103/PhysRevA.73.062501} {\bibfield  {journal} {\bibinfo  {journal} {Phys. Rev. A}\ }\textbf {\bibinfo {volume} {73}},\ \bibinfo {pages} {062501} (\bibinfo {year} {2006})}\BibitemShut {NoStop}%
\bibitem [{\citenamefont {Garcia~Ruiz}\ \emph {et~al.}(2018)\citenamefont {Garcia~Ruiz}, \citenamefont {Vernon}, \citenamefont {Binnersley}, \citenamefont {Sahoo}, \citenamefont {Bissell}, \citenamefont {Billowes}, \citenamefont {Cocolios}, \citenamefont {Gins}, \citenamefont {de~Groote}, \citenamefont {Flanagan}, \citenamefont {Koszorus}, \citenamefont {Lynch}, \citenamefont {Neyens}, \citenamefont {Ricketts}, \citenamefont {Wendt}, \citenamefont {Wilkins},\ and\ \citenamefont {Yang}}]{bijaya-in}%
  \BibitemOpen
  \bibfield  {author} {\bibinfo {author} {\bibfnamefont {R.~F.}\ \bibnamefont {Garcia~Ruiz}}, \bibinfo {author} {\bibfnamefont {A.~R.}\ \bibnamefont {Vernon}}, \bibinfo {author} {\bibfnamefont {C.~L.}\ \bibnamefont {Binnersley}}, \bibinfo {author} {\bibfnamefont {B.~K.}\ \bibnamefont {Sahoo}}, \bibinfo {author} {\bibfnamefont {M.}~\bibnamefont {Bissell}}, \bibinfo {author} {\bibfnamefont {J.}~\bibnamefont {Billowes}}, \bibinfo {author} {\bibfnamefont {T.~E.}\ \bibnamefont {Cocolios}}, \bibinfo {author} {\bibfnamefont {W.}~\bibnamefont {Gins}}, \bibinfo {author} {\bibfnamefont {R.~P.}\ \bibnamefont {de~Groote}}, \bibinfo {author} {\bibfnamefont {K.~T.}\ \bibnamefont {Flanagan}}, \bibinfo {author} {\bibfnamefont {A.}~\bibnamefont {Koszorus}}, \bibinfo {author} {\bibfnamefont {K.~M.}\ \bibnamefont {Lynch}}, \bibinfo {author} {\bibfnamefont {G.}~\bibnamefont {Neyens}}, \bibinfo {author} {\bibfnamefont {C.~M.}\ \bibnamefont {Ricketts}}, \bibinfo {author} {\bibfnamefont {K.~D.~A.}\ \bibnamefont {Wendt}}, \bibinfo
  {author} {\bibfnamefont {S.~G.}\ \bibnamefont {Wilkins}},\ and\ \bibinfo {author} {\bibfnamefont {X.~F.}\ \bibnamefont {Yang}},\ }\bibfield  {title} {\bibinfo {title} {High-precision multiphoton ionization of accelerated laser-ablated species},\ }\href {https://doi.org/10.1103/PhysRevX.8.041005} {\bibfield  {journal} {\bibinfo  {journal} {Phys. Rev. X}\ }\textbf {\bibinfo {volume} {8}},\ \bibinfo {pages} {041005} (\bibinfo {year} {2018})}\BibitemShut {NoStop}%
\bibitem [{\citenamefont {Das}\ \emph {et~al.}(2005)\citenamefont {Das}, \citenamefont {Latha}, \citenamefont {Sahoo}, \citenamefont {Sur}, \citenamefont {Chaudhuri},\ and\ \citenamefont {Mukherjee}}]{das}%
  \BibitemOpen
  \bibfield  {author} {\bibinfo {author} {\bibfnamefont {B.~P.}\ \bibnamefont {Das}}, \bibinfo {author} {\bibfnamefont {K.~V.~P.}\ \bibnamefont {Latha}}, \bibinfo {author} {\bibfnamefont {B.~K.}\ \bibnamefont {Sahoo}}, \bibinfo {author} {\bibfnamefont {C.}~\bibnamefont {Sur}}, \bibinfo {author} {\bibfnamefont {R.~K.}\ \bibnamefont {Chaudhuri}},\ and\ \bibinfo {author} {\bibfnamefont {D.}~\bibnamefont {Mukherjee}},\ }\bibfield  {title} {\bibinfo {title} {Relativistic and correlation effects in atoms},\ }\href {https://doi.org/10.1142/S0219633605001441} {\bibfield  {journal} {\bibinfo  {journal} {J. Theo. Comp. Chem.}\ }\textbf {\bibinfo {volume} {04}},\ \bibinfo {pages} {1} (\bibinfo {year} {2005})}\BibitemShut {NoStop}%
\bibitem [{\citenamefont {Grant}(2006)}]{grant2007relativistic}%
  \BibitemOpen
  \bibfield  {author} {\bibinfo {author} {\bibfnamefont {I.~P.}\ \bibnamefont {Grant}},\ }\href {https://doi.org/https://doi.org/10.1007/978-0-387-35069-1} {\emph {\bibinfo {title} {Relativistic Quantum Theory of Atoms and Molecules: Theory and Computation}}}\ (\bibinfo  {publisher} {SPRINGER NATURE},\ \bibinfo {year} {2006})\BibitemShut {NoStop}%
\bibitem [{\citenamefont {Jönsson}\ \emph {et~al.}(2023)\citenamefont {Jönsson}, \citenamefont {Godefroid}, \citenamefont {Gaigalas}, \citenamefont {Ekman}, \citenamefont {Grumer}, \citenamefont {Li}, \citenamefont {Li}, \citenamefont {Brage}, \citenamefont {Grant}, \citenamefont {Bieroń},\ and\ \citenamefont {Fischer}}]{atoms11010007}%
  \BibitemOpen
  \bibfield  {author} {\bibinfo {author} {\bibfnamefont {P.}~\bibnamefont {Jönsson}}, \bibinfo {author} {\bibfnamefont {M.}~\bibnamefont {Godefroid}}, \bibinfo {author} {\bibfnamefont {G.}~\bibnamefont {Gaigalas}}, \bibinfo {author} {\bibfnamefont {J.}~\bibnamefont {Ekman}}, \bibinfo {author} {\bibfnamefont {J.}~\bibnamefont {Grumer}}, \bibinfo {author} {\bibfnamefont {W.}~\bibnamefont {Li}}, \bibinfo {author} {\bibfnamefont {J.}~\bibnamefont {Li}}, \bibinfo {author} {\bibfnamefont {T.}~\bibnamefont {Brage}}, \bibinfo {author} {\bibfnamefont {I.~P.}\ \bibnamefont {Grant}}, \bibinfo {author} {\bibfnamefont {J.}~\bibnamefont {Bieroń}},\ and\ \bibinfo {author} {\bibfnamefont {C.~F.}\ \bibnamefont {Fischer}},\ }\bibfield  {title} {\bibinfo {title} {An introduction to relativistic theory as implemented in grasp},\ }\bibfield  {journal} {\bibinfo  {journal} {Atoms}\ }\textbf {\bibinfo {volume} {11}},\ \href {https://doi.org/10.3390/atoms11010007} {10.3390/atoms11010007} (\bibinfo {year} {2023})\BibitemShut
  {NoStop}%
\bibitem [{\citenamefont {Kumar}\ \emph {et~al.}(2018)\citenamefont {Kumar}, \citenamefont {Li},\ and\ \citenamefont {Sahoo}}]{bijaya-pradeep}%
  \BibitemOpen
  \bibfield  {author} {\bibinfo {author} {\bibfnamefont {P.}~\bibnamefont {Kumar}}, \bibinfo {author} {\bibfnamefont {C.-B.}\ \bibnamefont {Li}},\ and\ \bibinfo {author} {\bibfnamefont {B.~K.}\ \bibnamefont {Sahoo}},\ }\bibfield  {title} {\bibinfo {title} {Diverse trends of electron correlation effects for properties with different radial and angular factors in an atomic system: a case study in ca+},\ }\href {https://doi.org/10.1088/1361-6455/aaaa12} {\bibfield  {journal} {\bibinfo  {journal} {J. Phys. B: Atomic, Molecular and Optical Physics}\ }\textbf {\bibinfo {volume} {51}},\ \bibinfo {pages} {055101} (\bibinfo {year} {2018})}\BibitemShut {NoStop}%
\bibitem [{\citenamefont {Sahoo}\ \emph {et~al.}(2006)\citenamefont {Sahoo}, \citenamefont {Islam}, \citenamefont {Das}, \citenamefont {Chaudhuri},\ and\ \citenamefont {Mukherjee}}]{Lifetime}%
  \BibitemOpen
  \bibfield  {author} {\bibinfo {author} {\bibfnamefont {B.~K.}\ \bibnamefont {Sahoo}}, \bibinfo {author} {\bibfnamefont {M.~R.}\ \bibnamefont {Islam}}, \bibinfo {author} {\bibfnamefont {B.~P.}\ \bibnamefont {Das}}, \bibinfo {author} {\bibfnamefont {R.~K.}\ \bibnamefont {Chaudhuri}},\ and\ \bibinfo {author} {\bibfnamefont {D.}~\bibnamefont {Mukherjee}},\ }\bibfield  {title} {\bibinfo {title} {Lifetimes of the metastable $^{2}d_{3/2,5/2}$ states in ${\mathrm{ca}}^{+}$, ${\mathrm{sr}}^{+}$, and ${\mathrm{ba}}^{+}$},\ }\href {https://doi.org/10.1103/PhysRevA.74.062504} {\bibfield  {journal} {\bibinfo  {journal} {Phys. Rev. A}\ }\textbf {\bibinfo {volume} {74}},\ \bibinfo {pages} {062504} (\bibinfo {year} {2006})}\BibitemShut {NoStop}%
\bibitem [{\citenamefont {Sahoo}\ and\ \citenamefont {Das}(2008)}]{polz}%
  \BibitemOpen
  \bibfield  {author} {\bibinfo {author} {\bibfnamefont {B.~K.}\ \bibnamefont {Sahoo}}\ and\ \bibinfo {author} {\bibfnamefont {B.~P.}\ \bibnamefont {Das}},\ }\bibfield  {title} {\bibinfo {title} {Relativistic coupled-cluster studies of dipole polarizabilities in closed-shell atoms},\ }\href {https://doi.org/10.1103/PhysRevA.77.062516} {\bibfield  {journal} {\bibinfo  {journal} {Phys. Rev. A}\ }\textbf {\bibinfo {volume} {77}},\ \bibinfo {pages} {062516} (\bibinfo {year} {2008})}\BibitemShut {NoStop}%
\bibitem [{\citenamefont {Skripnikov}\ \emph {et~al.}(2024)\citenamefont {Skripnikov}, \citenamefont {Prosnyak}, \citenamefont {Malyshev}, \citenamefont {Athanasakis-Kaklamanakis}, \citenamefont {Brinson}, \citenamefont {Minamisono}, \citenamefont {Cruz}, \citenamefont {Reilly}, \citenamefont {Rickey},\ and\ \citenamefont {Ruiz}}]{Leonid}%
  \BibitemOpen
  \bibfield  {author} {\bibinfo {author} {\bibfnamefont {L.~V.}\ \bibnamefont {Skripnikov}}, \bibinfo {author} {\bibfnamefont {S.~D.}\ \bibnamefont {Prosnyak}}, \bibinfo {author} {\bibfnamefont {A.~V.}\ \bibnamefont {Malyshev}}, \bibinfo {author} {\bibfnamefont {M.}~\bibnamefont {Athanasakis-Kaklamanakis}}, \bibinfo {author} {\bibfnamefont {A.~J.}\ \bibnamefont {Brinson}}, \bibinfo {author} {\bibfnamefont {K.}~\bibnamefont {Minamisono}}, \bibinfo {author} {\bibfnamefont {F.~C.~P.}\ \bibnamefont {Cruz}}, \bibinfo {author} {\bibfnamefont {J.~R.}\ \bibnamefont {Reilly}}, \bibinfo {author} {\bibfnamefont {B.~J.}\ \bibnamefont {Rickey}},\ and\ \bibinfo {author} {\bibfnamefont {R.~F.~G.}\ \bibnamefont {Ruiz}},\ }\bibfield  {title} {\bibinfo {title} {Isotope-shift factors with quantum electrodynamics effects for many-electron systems: A study of the nuclear charge radius of $^{26m}\mathrm{Al}$},\ }\href {https://doi.org/10.1103/PhysRevA.110.012807} {\bibfield  {journal} {\bibinfo  {journal} {Phys. Rev. A}\ }\textbf
  {\bibinfo {volume} {110}},\ \bibinfo {pages} {012807} (\bibinfo {year} {2024})}\BibitemShut {NoStop}%
\bibitem [{\citenamefont {Shavitt}\ and\ \citenamefont {Bartlett}(2009)}]{bartlett}%
  \BibitemOpen
  \bibfield  {author} {\bibinfo {author} {\bibfnamefont {I.}~\bibnamefont {Shavitt}}\ and\ \bibinfo {author} {\bibfnamefont {R.~J.}\ \bibnamefont {Bartlett}},\ }\href {https://doi.org/https://doi.org/10.1017/CBO9780511596834} {\emph {\bibinfo {title} {Many-body methods in Chemistry and Physics}}}\ (\bibinfo  {publisher} {Cambidge University Press},\ \bibinfo {address} {Cambridge, UK},\ \bibinfo {year} {2009})\BibitemShut {NoStop}%
\bibitem [{\citenamefont {DeYonker}\ and\ \citenamefont {Peterson}(2013)}]{deyonker}%
  \BibitemOpen
  \bibfield  {author} {\bibinfo {author} {\bibfnamefont {N.~J.}\ \bibnamefont {DeYonker}}\ and\ \bibinfo {author} {\bibfnamefont {K.~A.}\ \bibnamefont {Peterson}},\ }\bibfield  {title} {\bibinfo {title} {Is near-“spectroscopic accuracy” possible for heavy atoms and coupled cluster theory? an investigation of the first ionization potentials of the atoms ga–kr},\ }\href {https://doi.org/10.1063/1.4801854} {\bibfield  {journal} {\bibinfo  {journal} {J. Chem. Phys.}\ }\textbf {\bibinfo {volume} {138}},\ \bibinfo {pages} {164312} (\bibinfo {year} {2013})}\BibitemShut {NoStop}%
\bibitem [{\citenamefont {Nataraj}\ \emph {et~al.}(2008{\natexlab{b}})\citenamefont {Nataraj}, \citenamefont {Sahoo}, \citenamefont {Das},\ and\ \citenamefont {Mukherjee}}]{nataraj}%
  \BibitemOpen
  \bibfield  {author} {\bibinfo {author} {\bibfnamefont {H.~S.}\ \bibnamefont {Nataraj}}, \bibinfo {author} {\bibfnamefont {B.~K.}\ \bibnamefont {Sahoo}}, \bibinfo {author} {\bibfnamefont {B.~P.}\ \bibnamefont {Das}},\ and\ \bibinfo {author} {\bibfnamefont {D.}~\bibnamefont {Mukherjee}},\ }\bibfield  {title} {\bibinfo {title} {Intrinsic electric dipole moments of paramagnetic atoms: Rubidium and cesium},\ }\href {https://doi.org/10.1103/PhysRevLett.101.033002} {\bibfield  {journal} {\bibinfo  {journal} {Phys. Rev. Lett.}\ }\textbf {\bibinfo {volume} {101}},\ \bibinfo {pages} {033002} (\bibinfo {year} {2008}{\natexlab{b}})}\BibitemShut {NoStop}%
\bibitem [{\citenamefont {Pa\ifmmode~\check{s}\else \v{s}\fi{}teka}\ \emph {et~al.}(2017)\citenamefont {Pa\ifmmode~\check{s}\else \v{s}\fi{}teka}, \citenamefont {Eliav}, \citenamefont {Borschevsky}, \citenamefont {Kaldor},\ and\ \citenamefont {Schwerdtfeger}}]{eliav}%
  \BibitemOpen
  \bibfield  {author} {\bibinfo {author} {\bibfnamefont {L.~F.}\ \bibnamefont {Pa\ifmmode~\check{s}\else \v{s}\fi{}teka}}, \bibinfo {author} {\bibfnamefont {E.}~\bibnamefont {Eliav}}, \bibinfo {author} {\bibfnamefont {A.}~\bibnamefont {Borschevsky}}, \bibinfo {author} {\bibfnamefont {U.}~\bibnamefont {Kaldor}},\ and\ \bibinfo {author} {\bibfnamefont {P.}~\bibnamefont {Schwerdtfeger}},\ }\bibfield  {title} {\bibinfo {title} {Relativistic coupled cluster calculations with variational quantum electrodynamics resolve the discrepancy between experiment and theory concerning the electron affinity and ionization potential of gold},\ }\href {https://doi.org/10.1103/PhysRevLett.118.023002} {\bibfield  {journal} {\bibinfo  {journal} {Phys. Rev. Lett.}\ }\textbf {\bibinfo {volume} {118}},\ \bibinfo {pages} {023002} (\bibinfo {year} {2017})}\BibitemShut {NoStop}%
\bibitem [{\citenamefont {Čížek}(1969)}]{cizek}%
  \BibitemOpen
  \bibfield  {author} {\bibinfo {author} {\bibfnamefont {J.}~\bibnamefont {Čížek}},\ }\bibinfo {title} {On the use of the cluster expansion and the technique of diagrams in calculations of correlation effects in atoms and molecules},\ in\ \href {https://doi.org/https://doi.org/10.1002/9780470143599.ch2} {\emph {\bibinfo {booktitle} {Adv. Chem. Phys.}}}\ (\bibinfo  {publisher} {John Wiley \& Sons, Ltd},\ \bibinfo {year} {1969})\ pp.\ \bibinfo {pages} {35--89}\BibitemShut {NoStop}%
\bibitem [{\citenamefont {Crawford}\ and\ \citenamefont {Schaefer~III}(2000)}]{crawford}%
  \BibitemOpen
  \bibfield  {author} {\bibinfo {author} {\bibfnamefont {T.~D.}\ \bibnamefont {Crawford}}\ and\ \bibinfo {author} {\bibfnamefont {H.~F.}\ \bibnamefont {Schaefer~III}},\ }\bibinfo {title} {An introduction to coupled cluster theory for computational chemists},\ in\ \href {https://doi.org/https://doi.org/10.1002/9780470125915.ch2} {\emph {\bibinfo {booktitle} {Rev. Comp. Chem.}}}\ (\bibinfo  {publisher} {John Wiley \& Sons, Ltd},\ \bibinfo {year} {2000})\ pp.\ \bibinfo {pages} {33--136}\BibitemShut {NoStop}%
\bibitem [{\citenamefont {Prasannaa}\ \emph {et~al.}(2015)\citenamefont {Prasannaa}, \citenamefont {Vutha}, \citenamefont {Abe},\ and\ \citenamefont {Das}}]{prasana}%
  \BibitemOpen
  \bibfield  {author} {\bibinfo {author} {\bibfnamefont {V.~S.}\ \bibnamefont {Prasannaa}}, \bibinfo {author} {\bibfnamefont {A.~C.}\ \bibnamefont {Vutha}}, \bibinfo {author} {\bibfnamefont {M.}~\bibnamefont {Abe}},\ and\ \bibinfo {author} {\bibfnamefont {B.~P.}\ \bibnamefont {Das}},\ }\bibfield  {title} {\bibinfo {title} {Mercury monohalides: Suitability for electron electric dipole moment searches},\ }\href {https://doi.org/10.1103/PhysRevLett.114.183001} {\bibfield  {journal} {\bibinfo  {journal} {Phys. Rev. Lett.}\ }\textbf {\bibinfo {volume} {114}},\ \bibinfo {pages} {183001} (\bibinfo {year} {2015})}\BibitemShut {NoStop}%
\bibitem [{\citenamefont {Bishop}\ and\ \citenamefont {Li}(2017)}]{bishop01}%
  \BibitemOpen
  \bibfield  {author} {\bibinfo {author} {\bibfnamefont {R.~F.}\ \bibnamefont {Bishop}}\ and\ \bibinfo {author} {\bibfnamefont {P.~H.~Y.}\ \bibnamefont {Li}},\ }\bibfield  {title} {\bibinfo {title} {Frustrated honeycomb-bilayer heisenberg antiferromagnet: The spin-$\frac{1}{2}$ ${J}_{1}\text{\ensuremath{-}}{J}_{2}\text{\ensuremath{-}}{J}_{1}^{\ensuremath{\perp}}$ model},\ }\href {https://doi.org/10.1103/PhysRevB.95.134414} {\bibfield  {journal} {\bibinfo  {journal} {Phys. Rev. B}\ }\textbf {\bibinfo {volume} {95}},\ \bibinfo {pages} {134414} (\bibinfo {year} {2017})}\BibitemShut {NoStop}%
\bibitem [{\citenamefont {Bishop}(1991)}]{bishop}%
  \BibitemOpen
  \bibfield  {author} {\bibinfo {author} {\bibfnamefont {R.~F.}\ \bibnamefont {Bishop}},\ }\bibfield  {title} {\bibinfo {title} {An overview of coupled cluster theory and its applications in physics},\ }\href {https://doi.org/10.1007/BF01119617} {\bibfield  {journal} {\bibinfo  {journal} {Theor. Chim. Acta}\ }\textbf {\bibinfo {volume} {80}},\ \bibinfo {pages} {95} (\bibinfo {year} {1991})}\BibitemShut {NoStop}%
\bibitem [{\citenamefont {Kowalski}\ \emph {et~al.}(2004)\citenamefont {Kowalski}, \citenamefont {Dean}, \citenamefont {Hjorth-Jensen}, \citenamefont {Papenbrock},\ and\ \citenamefont {Piecuch}}]{kowalski}%
  \BibitemOpen
  \bibfield  {author} {\bibinfo {author} {\bibfnamefont {K.}~\bibnamefont {Kowalski}}, \bibinfo {author} {\bibfnamefont {D.~J.}\ \bibnamefont {Dean}}, \bibinfo {author} {\bibfnamefont {M.}~\bibnamefont {Hjorth-Jensen}}, \bibinfo {author} {\bibfnamefont {T.}~\bibnamefont {Papenbrock}},\ and\ \bibinfo {author} {\bibfnamefont {P.}~\bibnamefont {Piecuch}},\ }\bibfield  {title} {\bibinfo {title} {Coupled cluster calculations of ground and excited states of nuclei},\ }\href {https://doi.org/10.1103/PhysRevLett.92.132501} {\bibfield  {journal} {\bibinfo  {journal} {Phys. Rev. Lett.}\ }\textbf {\bibinfo {volume} {92}},\ \bibinfo {pages} {132501} (\bibinfo {year} {2004})}\BibitemShut {NoStop}%
\bibitem [{\citenamefont {Hagen}\ \emph {et~al.}(2014)\citenamefont {Hagen}, \citenamefont {Papenbrock}, \citenamefont {Hjorth-Jensen},\ and\ \citenamefont {Dean}}]{hagen}%
  \BibitemOpen
  \bibfield  {author} {\bibinfo {author} {\bibfnamefont {G.}~\bibnamefont {Hagen}}, \bibinfo {author} {\bibfnamefont {T.}~\bibnamefont {Papenbrock}}, \bibinfo {author} {\bibfnamefont {M.}~\bibnamefont {Hjorth-Jensen}},\ and\ \bibinfo {author} {\bibfnamefont {D.~J.}\ \bibnamefont {Dean}},\ }\bibfield  {title} {\bibinfo {title} {Coupled-cluster computations of atomic nuclei},\ }\href {https://doi.org/10.1088/0034-4885/77/9/096302} {\bibfield  {journal} {\bibinfo  {journal} {Rep. Prog. Phys.}\ }\textbf {\bibinfo {volume} {77}},\ \bibinfo {pages} {096302} (\bibinfo {year} {2014})}\BibitemShut {NoStop}%
\bibitem [{\citenamefont {Bishop}\ \emph {et~al.}(1989)\citenamefont {Bishop}, \citenamefont {Arponen},\ and\ \citenamefont {Pajanee}}]{bishopbook1}%
  \BibitemOpen
  \bibfield  {author} {\bibinfo {author} {\bibfnamefont {R.}~\bibnamefont {Bishop}}, \bibinfo {author} {\bibfnamefont {J.}~\bibnamefont {Arponen}},\ and\ \bibinfo {author} {\bibfnamefont {P.}~\bibnamefont {Pajanee}},\ }\href {https://doi.org/https://doi.org/10.1007/978-3-642-61330-2} {\emph {\bibinfo {title} {Aspects of Many-body Effects in Molecules and Extended Systems}}}\ (\bibinfo  {publisher} {Springer-Verlag},\ \bibinfo {address} {Berlin},\ \bibinfo {year} {1989})\BibitemShut {NoStop}%
\bibitem [{\citenamefont {Bishop}(1998)}]{bishopbook}%
  \BibitemOpen
  \bibfield  {author} {\bibinfo {author} {\bibfnamefont {R.~F.}\ \bibnamefont {Bishop}},\ }\bibfield  {title} {\bibinfo {title} {The coupled cluster method},\ }in\ \href {https://doi.org/https://doi.org/10.1007/BFb0104523} {\emph {\bibinfo {booktitle} {Microscopic Quantum Many-Body Theories and Their Applications}}},\ \bibinfo {editor} {edited by\ \bibinfo {editor} {\bibfnamefont {J.}~\bibnamefont {Navarro}}\ and\ \bibinfo {editor} {\bibfnamefont {A.}~\bibnamefont {Polls}}}\ (\bibinfo  {publisher} {Springer Berlin Heidelberg},\ \bibinfo {address} {Berlin, Heidelberg},\ \bibinfo {year} {1998})\ pp.\ \bibinfo {pages} {1--70}\BibitemShut {NoStop}%
\bibitem [{\citenamefont {Sahoo}\ and\ \citenamefont {Das}(2018)}]{bijaya-ncc}%
  \BibitemOpen
  \bibfield  {author} {\bibinfo {author} {\bibfnamefont {B.~K.}\ \bibnamefont {Sahoo}}\ and\ \bibinfo {author} {\bibfnamefont {B.~P.}\ \bibnamefont {Das}},\ }\bibfield  {title} {\bibinfo {title} {Relativistic normal coupled-cluster theory for accurate determination of electric dipole moments of atoms: First application to the $^{199}\mathrm{Hg}$ atom},\ }\href {https://doi.org/10.1103/PhysRevLett.120.203001} {\bibfield  {journal} {\bibinfo  {journal} {Phys. Rev. Lett.}\ }\textbf {\bibinfo {volume} {120}},\ \bibinfo {pages} {203001} (\bibinfo {year} {2018})}\BibitemShut {NoStop}%
\bibitem [{\citenamefont {Sahoo}\ \emph {et~al.}(2020)\citenamefont {Sahoo}, \citenamefont {Vernon}, \citenamefont {Garcia~Ruiz}, \citenamefont {Binnersley}, \citenamefont {Billowes}, \citenamefont {Bissell}, \citenamefont {Cocolios}, \citenamefont {Farooq-Smith}, \citenamefont {Flanagan}, \citenamefont {Gins}, \citenamefont {de~Groote}, \citenamefont {Koszorús}, \citenamefont {Neyens}, \citenamefont {Lynch}, \citenamefont {Parnefjord-Gustafsson}, \citenamefont {Ricketts}, \citenamefont {Wendt}, \citenamefont {Wilkins},\ and\ \citenamefont {Yang}}]{bijaya-ar}%
  \BibitemOpen
  \bibfield  {author} {\bibinfo {author} {\bibfnamefont {B.~K.}\ \bibnamefont {Sahoo}}, \bibinfo {author} {\bibfnamefont {A.~R.}\ \bibnamefont {Vernon}}, \bibinfo {author} {\bibfnamefont {R.~F.}\ \bibnamefont {Garcia~Ruiz}}, \bibinfo {author} {\bibfnamefont {C.~L.}\ \bibnamefont {Binnersley}}, \bibinfo {author} {\bibfnamefont {J.}~\bibnamefont {Billowes}}, \bibinfo {author} {\bibfnamefont {M.~L.}\ \bibnamefont {Bissell}}, \bibinfo {author} {\bibfnamefont {T.~E.}\ \bibnamefont {Cocolios}}, \bibinfo {author} {\bibfnamefont {G.~J.}\ \bibnamefont {Farooq-Smith}}, \bibinfo {author} {\bibfnamefont {K.~T.}\ \bibnamefont {Flanagan}}, \bibinfo {author} {\bibfnamefont {W.}~\bibnamefont {Gins}}, \bibinfo {author} {\bibfnamefont {R.~P.}\ \bibnamefont {de~Groote}}, \bibinfo {author} {\bibfnamefont {A.}~\bibnamefont {Koszorús}}, \bibinfo {author} {\bibfnamefont {G.}~\bibnamefont {Neyens}}, \bibinfo {author} {\bibfnamefont {K.~M.}\ \bibnamefont {Lynch}}, \bibinfo {author} {\bibfnamefont {F.}~\bibnamefont
  {Parnefjord-Gustafsson}}, \bibinfo {author} {\bibfnamefont {C.~M.}\ \bibnamefont {Ricketts}}, \bibinfo {author} {\bibfnamefont {K.~D.~A.}\ \bibnamefont {Wendt}}, \bibinfo {author} {\bibfnamefont {S.~G.}\ \bibnamefont {Wilkins}},\ and\ \bibinfo {author} {\bibfnamefont {X.~F.}\ \bibnamefont {Yang}},\ }\bibfield  {title} {\bibinfo {title} {Analytic response relativistic coupled-cluster theory: the first application to indium isotope shifts},\ }\href {https://doi.org/10.1088/1367-2630/ab66dd} {\bibfield  {journal} {\bibinfo  {journal} {New J. Phys.}\ }\textbf {\bibinfo {volume} {22}},\ \bibinfo {pages} {012001} (\bibinfo {year} {2020})}\BibitemShut {NoStop}%
\bibitem [{\citenamefont {Katyal}\ \emph {et~al.}(2025)\citenamefont {Katyal}, \citenamefont {Chakraborty}, \citenamefont {Sahoo}, \citenamefont {Ohayon}, \citenamefont {Seng}, \citenamefont {Gorchtein},\ and\ \citenamefont {Behr}}]{bijaya-k}%
  \BibitemOpen
  \bibfield  {author} {\bibinfo {author} {\bibfnamefont {V.}~\bibnamefont {Katyal}}, \bibinfo {author} {\bibfnamefont {A.}~\bibnamefont {Chakraborty}}, \bibinfo {author} {\bibfnamefont {B.~K.}\ \bibnamefont {Sahoo}}, \bibinfo {author} {\bibfnamefont {B.}~\bibnamefont {Ohayon}}, \bibinfo {author} {\bibfnamefont {C.-Y.}\ \bibnamefont {Seng}}, \bibinfo {author} {\bibfnamefont {M.}~\bibnamefont {Gorchtein}},\ and\ \bibinfo {author} {\bibfnamefont {J.}~\bibnamefont {Behr}},\ }\bibfield  {title} {\bibinfo {title} {Testing for isospin symmetry breaking by combining isotope shift measurements with precise calculations in potassium},\ }\href {https://doi.org/10.1103/PhysRevA.111.042813} {\bibfield  {journal} {\bibinfo  {journal} {Phys. Rev. A}\ }\textbf {\bibinfo {volume} {111}},\ \bibinfo {pages} {042813} (\bibinfo {year} {2025})}\BibitemShut {NoStop}%
\bibitem [{\citenamefont {Jönsson}\ \emph {et~al.}(2017)\citenamefont {Jönsson}, \citenamefont {Gaigalas}, \citenamefont {Rynkun}, \citenamefont {Radžiūtė}, \citenamefont {Ekman}, \citenamefont {Gustafsson}, \citenamefont {Hartman}, \citenamefont {Wang}, \citenamefont {Godefroid}, \citenamefont {Froese~Fischer}, \citenamefont {Grant}, \citenamefont {Brage},\ and\ \citenamefont {Del~Zanna}}]{atoms5020016}%
  \BibitemOpen
  \bibfield  {author} {\bibinfo {author} {\bibfnamefont {P.}~\bibnamefont {Jönsson}}, \bibinfo {author} {\bibfnamefont {G.}~\bibnamefont {Gaigalas}}, \bibinfo {author} {\bibfnamefont {P.}~\bibnamefont {Rynkun}}, \bibinfo {author} {\bibfnamefont {L.}~\bibnamefont {Radžiūtė}}, \bibinfo {author} {\bibfnamefont {J.}~\bibnamefont {Ekman}}, \bibinfo {author} {\bibfnamefont {S.}~\bibnamefont {Gustafsson}}, \bibinfo {author} {\bibfnamefont {H.}~\bibnamefont {Hartman}}, \bibinfo {author} {\bibfnamefont {K.}~\bibnamefont {Wang}}, \bibinfo {author} {\bibfnamefont {M.}~\bibnamefont {Godefroid}}, \bibinfo {author} {\bibfnamefont {C.}~\bibnamefont {Froese~Fischer}}, \bibinfo {author} {\bibfnamefont {I.}~\bibnamefont {Grant}}, \bibinfo {author} {\bibfnamefont {T.}~\bibnamefont {Brage}},\ and\ \bibinfo {author} {\bibfnamefont {G.}~\bibnamefont {Del~Zanna}},\ }\bibfield  {title} {\bibinfo {title} {Multiconfiguration dirac-hartree-fock calculations with spectroscopic accuracy: Applications to astrophysics},\ }\bibfield
  {journal} {\bibinfo  {journal} {Atoms}\ }\textbf {\bibinfo {volume} {5}},\ \href {https://doi.org/10.3390/atoms5020016} {10.3390/atoms5020016} (\bibinfo {year} {2017})\BibitemShut {NoStop}%
\bibitem [{\citenamefont {Lindgren}\ and\ \citenamefont {Morrison}(1986)}]{Lindgren1986}%
  \BibitemOpen
  \bibfield  {author} {\bibinfo {author} {\bibfnamefont {I.}~\bibnamefont {Lindgren}}\ and\ \bibinfo {author} {\bibfnamefont {J.}~\bibnamefont {Morrison}},\ }\href {https://doi.org/https://doi.org/10.1007/978-3-642-61640-2} {\emph {\bibinfo {title} {Atomic Many-body Theory}}}\ (\bibinfo  {publisher} {Springer-Verlag},\ \bibinfo {address} {Berlin, Heidelberg, New York},\ \bibinfo {year} {1986})\BibitemShut {NoStop}%
\bibitem [{\citenamefont {Kramida}\ \emph {et~al.}(2024)\citenamefont {Kramida}, \citenamefont {{Yu.~Ralchenko}}, \citenamefont {Reader},\ and\ \citenamefont {{and NIST ASD Team}}}]{NISTdata}%
  \BibitemOpen
  \bibfield  {author} {\bibinfo {author} {\bibfnamefont {A.}~\bibnamefont {Kramida}}, \bibinfo {author} {\bibnamefont {{Yu.~Ralchenko}}}, \bibinfo {author} {\bibfnamefont {J.}~\bibnamefont {Reader}},\ and\ \bibinfo {author} {\bibnamefont {{and NIST ASD Team}}},\ }\href {https://doi.org/https://dx.doi.org/10.18434/T4W30F} {\bibinfo {title} {Nist atomic spectra database (ver. 5.12)}} (\bibinfo {year} {2024})\BibitemShut {NoStop}%
\bibitem [{\citenamefont {Ekman}\ \emph {et~al.}(2019)\citenamefont {Ekman}, \citenamefont {Jönsson}, \citenamefont {Godefroid}, \citenamefont {Nazé}, \citenamefont {Gaigalas},\ and\ \citenamefont {Bieroń}}]{EKMAN2019433}%
  \BibitemOpen
  \bibfield  {author} {\bibinfo {author} {\bibfnamefont {J.}~\bibnamefont {Ekman}}, \bibinfo {author} {\bibfnamefont {P.}~\bibnamefont {Jönsson}}, \bibinfo {author} {\bibfnamefont {M.}~\bibnamefont {Godefroid}}, \bibinfo {author} {\bibfnamefont {C.}~\bibnamefont {Nazé}}, \bibinfo {author} {\bibfnamefont {G.}~\bibnamefont {Gaigalas}},\ and\ \bibinfo {author} {\bibfnamefont {J.}~\bibnamefont {Bieroń}},\ }\bibfield  {title} {\bibinfo {title} {ris 4: A program for relativistic isotope shift calculations},\ }\href {https://doi.org/https://doi.org/10.1016/j.cpc.2018.08.017} {\bibfield  {journal} {\bibinfo  {journal} {Comp. Phys. Comm.}\ }\textbf {\bibinfo {volume} {235}},\ \bibinfo {pages} {433} (\bibinfo {year} {2019})}\BibitemShut {NoStop}%
\bibitem [{\citenamefont {Gaidamauskas}\ \emph {et~al.}(2011)\citenamefont {Gaidamauskas}, \citenamefont {Naze}, \citenamefont {Rynkun}, \citenamefont {Gaigalas}, \citenamefont {Joensson},\ and\ \citenamefont {Godefroid}}]{Gaidamauskas}%
  \BibitemOpen
  \bibfield  {author} {\bibinfo {author} {\bibfnamefont {E.}~\bibnamefont {Gaidamauskas}}, \bibinfo {author} {\bibfnamefont {C.}~\bibnamefont {Naze}}, \bibinfo {author} {\bibfnamefont {P.}~\bibnamefont {Rynkun}}, \bibinfo {author} {\bibfnamefont {G.}~\bibnamefont {Gaigalas}}, \bibinfo {author} {\bibfnamefont {P.}~\bibnamefont {Joensson}},\ and\ \bibinfo {author} {\bibfnamefont {M.}~\bibnamefont {Godefroid}},\ }\bibfield  {title} {\bibinfo {title} {Tensorial form and matrix elements of the relativistic nuclear recoil operator},\ }\href {https://doi.org/10.1088/0953-4075/44/17/175003} {\bibfield  {journal} {\bibinfo  {journal} {J. Phys. B: Atomic, Molecular and Optical Physics}\ }\textbf {\bibinfo {volume} {44}},\ \bibinfo {pages} {175003} (\bibinfo {year} {2011})}\BibitemShut {NoStop}%
\bibitem [{\citenamefont {Fischer}\ \emph {et~al.}(2016)\citenamefont {Fischer}, \citenamefont {Godefroid}, \citenamefont {Brage}, \citenamefont {J{\"o}nsson},\ and\ \citenamefont {Gaigalas}}]{fischer2016advanced}%
  \BibitemOpen
  \bibfield  {author} {\bibinfo {author} {\bibfnamefont {C.~F.}\ \bibnamefont {Fischer}}, \bibinfo {author} {\bibfnamefont {M.}~\bibnamefont {Godefroid}}, \bibinfo {author} {\bibfnamefont {T.}~\bibnamefont {Brage}}, \bibinfo {author} {\bibfnamefont {P.}~\bibnamefont {J{\"o}nsson}},\ and\ \bibinfo {author} {\bibfnamefont {G.}~\bibnamefont {Gaigalas}},\ }\bibfield  {title} {\bibinfo {title} {Advanced multiconfiguration methods for complex atoms: I. energies and wave functions},\ }\href {https://doi.org/10.1088/0953-4075/49/18/182004} {\bibfield  {journal} {\bibinfo  {journal} {J. Phys. B: Atomic, Molecular and Optical Physics}\ }\textbf {\bibinfo {volume} {49}},\ \bibinfo {pages} {182004} (\bibinfo {year} {2016})}\BibitemShut {NoStop}%
\bibitem [{\citenamefont {Mackenzie}\ \emph {et~al.}(1980)\citenamefont {Mackenzie}, \citenamefont {Grant},\ and\ \citenamefont {Norrington}}]{mackenzie1980program}%
  \BibitemOpen
  \bibfield  {author} {\bibinfo {author} {\bibfnamefont {B.}~\bibnamefont {Mackenzie}}, \bibinfo {author} {\bibfnamefont {I.}~\bibnamefont {Grant}},\ and\ \bibinfo {author} {\bibfnamefont {P.}~\bibnamefont {Norrington}},\ }\bibfield  {title} {\bibinfo {title} {Program to calculate transverse breit and qed corrections to energy levels in a multiconfiguration dirac-fock environment},\ }\bibfield  {journal} {\bibinfo  {journal} {Comput. Phys. Commun.;(Netherlands)}\ }\textbf {\bibinfo {volume} {21}},\ \href {https://doi.org/10.1016/0010-4655(80)90042-9} {10.1016/0010-4655(80)90042-9} (\bibinfo {year} {1980})\BibitemShut {NoStop}%
\bibitem [{\citenamefont {Gaigalas}(2022)}]{atoms10040129}%
  \BibitemOpen
  \bibfield  {author} {\bibinfo {author} {\bibfnamefont {G.}~\bibnamefont {Gaigalas}},\ }\bibfield  {title} {\bibinfo {title} {A program library for computing pure spin–angular coefficients for one- and two-particle operators in relativistic atomic theory},\ }\bibfield  {journal} {\bibinfo  {journal} {Atoms}\ }\textbf {\bibinfo {volume} {10}},\ \href {https://doi.org/10.3390/atoms10040129} {10.3390/atoms10040129} (\bibinfo {year} {2022})\BibitemShut {NoStop}%
\bibitem [{\citenamefont {Li}\ \emph {et~al.}(2022)\citenamefont {Li}, \citenamefont {Wang}, \citenamefont {Si}, \citenamefont {Godefroid}, \citenamefont {Gaigalas}, \citenamefont {Chen},\ and\ \citenamefont {J{\"o}nsson}}]{YTLiCPC}%
  \BibitemOpen
  \bibfield  {author} {\bibinfo {author} {\bibfnamefont {Y.~T.}\ \bibnamefont {Li}}, \bibinfo {author} {\bibfnamefont {K.}~\bibnamefont {Wang}}, \bibinfo {author} {\bibfnamefont {R.}~\bibnamefont {Si}}, \bibinfo {author} {\bibfnamefont {M.}~\bibnamefont {Godefroid}}, \bibinfo {author} {\bibfnamefont {G.}~\bibnamefont {Gaigalas}}, \bibinfo {author} {\bibfnamefont {C.~Y.}\ \bibnamefont {Chen}},\ and\ \bibinfo {author} {\bibfnamefont {P.}~\bibnamefont {J{\"o}nsson}},\ }\bibfield  {title} {\bibinfo {title} {Reducing the computational load – atomic multiconfiguration calculations based on configuration state function generators},\ }\href {https://doi.org/10.1016/j.cpc.2022.108562} {\bibfield  {journal} {\bibinfo  {journal} {Comp. Phys. Comm.}\ ,\ \bibinfo {pages} {108562}} (\bibinfo {year} {2022})}\BibitemShut {NoStop}%
\bibitem [{\citenamefont {{Ansbacher}}\ \emph {et~al.}(1989)\citenamefont {{Ansbacher}}, \citenamefont {{Li}},\ and\ \citenamefont {{Pinnington}}}]{Ansbacher1989}%
  \BibitemOpen
  \bibfield  {author} {\bibinfo {author} {\bibfnamefont {W.}~\bibnamefont {{Ansbacher}}}, \bibinfo {author} {\bibfnamefont {Y.}~\bibnamefont {{Li}}},\ and\ \bibinfo {author} {\bibfnamefont {E.~H.}\ \bibnamefont {{Pinnington}}},\ }\bibfield  {title} {\bibinfo {title} {{Precision lifetime measurement for the 3p levels of Mg II using frequency-doubled laser radiation to excite a fast ion beam}},\ }\href {https://doi.org/10.1016/0375-9601(89)90353-8} {\bibfield  {journal} {\bibinfo  {journal} {Phys. Lett. A}\ }\textbf {\bibinfo {volume} {139}},\ \bibinfo {pages} {165} (\bibinfo {year} {1989})}\BibitemShut {NoStop}%
\bibitem [{\citenamefont {Lundin}\ \emph {et~al.}(1973)\citenamefont {Lundin}, \citenamefont {Engman}, \citenamefont {Hilke},\ and\ \citenamefont {Martinson}}]{Lundin1973}%
  \BibitemOpen
  \bibfield  {author} {\bibinfo {author} {\bibfnamefont {L.}~\bibnamefont {Lundin}}, \bibinfo {author} {\bibfnamefont {B.}~\bibnamefont {Engman}}, \bibinfo {author} {\bibfnamefont {J.}~\bibnamefont {Hilke}},\ and\ \bibinfo {author} {\bibfnamefont {I.}~\bibnamefont {Martinson}},\ }\bibfield  {title} {\bibinfo {title} {Lifetime measurements in mg i-mg iv},\ }\href {https://doi.org/10.1088/0031-8949/8/6/009} {\bibfield  {journal} {\bibinfo  {journal} {Phys. Scr.}\ }\textbf {\bibinfo {volume} {8}},\ \bibinfo {pages} {274} (\bibinfo {year} {1973})}\BibitemShut {NoStop}%
\bibitem [{\citenamefont {{Schaefer}}(1971)}]{Schaefer1971}%
  \BibitemOpen
  \bibfield  {author} {\bibinfo {author} {\bibfnamefont {A.~R.}\ \bibnamefont {{Schaefer}}},\ }\bibfield  {title} {\bibinfo {title} {{Measured Lifetimes of Excited States of Magnesium}},\ }\href {https://doi.org/10.1086/150781} {\bibfield  {journal} {\bibinfo  {journal} {\apj}\ }\textbf {\bibinfo {volume} {163}},\ \bibinfo {pages} {411} (\bibinfo {year} {1971})}\BibitemShut {NoStop}%
\bibitem [{\citenamefont {{Andersen}}(1970)}]{Andersen1970}%
  \BibitemOpen
  \bibfield  {author} {\bibinfo {author} {\bibfnamefont {T.}~\bibnamefont {{Andersen}}},\ }\bibfield  {title} {\bibinfo {title} {Measurements of atomic lifetimes for neutral and ionized magnesium and calcium},\ }\href {https://doi.org/10.1016/0022-4073(70)90063-4} {\bibfield  {journal} {\bibinfo  {journal} {J. Quant. Spectr. Rad. Trans.}\ }\textbf {\bibinfo {volume} {10}},\ \bibinfo {pages} {1143} (\bibinfo {year} {1970})}\BibitemShut {NoStop}%
\bibitem [{\citenamefont {Berry}\ \emph {et~al.}(1970)\citenamefont {Berry}, \citenamefont {Bromander},\ and\ \citenamefont {Buchta}}]{Berry1970}%
  \BibitemOpen
  \bibfield  {author} {\bibinfo {author} {\bibfnamefont {H.~G.}\ \bibnamefont {Berry}}, \bibinfo {author} {\bibfnamefont {J.}~\bibnamefont {Bromander}},\ and\ \bibinfo {author} {\bibfnamefont {R.}~\bibnamefont {Buchta}},\ }\bibfield  {title} {\bibinfo {title} {Some mean life measurements in the na i and mg i isoelectronic sequences},\ }\href {https://doi.org/10.1088/0031-8949/1/4/007} {\bibfield  {journal} {\bibinfo  {journal} {Phys. Scrip.}\ }\textbf {\bibinfo {volume} {1}},\ \bibinfo {pages} {181} (\bibinfo {year} {1970})}\BibitemShut {NoStop}%
\bibitem [{\citenamefont {Li}\ \emph {et~al.}(2020)\citenamefont {Li}, \citenamefont {Grumer}, \citenamefont {Brage},\ and\ \citenamefont {Jönsson}}]{LI2020107211}%
  \BibitemOpen
  \bibfield  {author} {\bibinfo {author} {\bibfnamefont {W.}~\bibnamefont {Li}}, \bibinfo {author} {\bibfnamefont {J.}~\bibnamefont {Grumer}}, \bibinfo {author} {\bibfnamefont {T.}~\bibnamefont {Brage}},\ and\ \bibinfo {author} {\bibfnamefont {P.}~\bibnamefont {Jönsson}},\ }\bibfield  {title} {\bibinfo {title} {Hfszeeman95—a program for computing weak and intermediate magnetic-field- and hyperfine-induced transition rates},\ }\href {https://doi.org/https://doi.org/10.1016/j.cpc.2020.107211} {\bibfield  {journal} {\bibinfo  {journal} {Comp. Phys. Comm.}\ }\textbf {\bibinfo {volume} {253}},\ \bibinfo {pages} {107211} (\bibinfo {year} {2020})}\BibitemShut {NoStop}%
\bibitem [{\citenamefont {Grant}(1974)}]{Grant1974}%
  \BibitemOpen
  \bibfield  {author} {\bibinfo {author} {\bibfnamefont {I.~P.}\ \bibnamefont {Grant}},\ }\bibfield  {title} {\bibinfo {title} {Gauge invariance and relativistic radiative transitions},\ }\href {https://doi.org/10.1088/0022-3700/7/12/007} {\bibfield  {journal} {\bibinfo  {journal} {J. Phys. B: Atomic and Molecular Physics}\ }\textbf {\bibinfo {volume} {7}},\ \bibinfo {pages} {1458} (\bibinfo {year} {1974})}\BibitemShut {NoStop}%
\bibitem [{\citenamefont {Dyall}\ \emph {et~al.}(1989)\citenamefont {Dyall}, \citenamefont {Grant}, \citenamefont {Johnson}, \citenamefont {Parpia},\ and\ \citenamefont {Plummer}}]{DYALL1989425}%
  \BibitemOpen
  \bibfield  {author} {\bibinfo {author} {\bibfnamefont {K.}~\bibnamefont {Dyall}}, \bibinfo {author} {\bibfnamefont {I.}~\bibnamefont {Grant}}, \bibinfo {author} {\bibfnamefont {C.}~\bibnamefont {Johnson}}, \bibinfo {author} {\bibfnamefont {F.}~\bibnamefont {Parpia}},\ and\ \bibinfo {author} {\bibfnamefont {E.}~\bibnamefont {Plummer}},\ }\bibfield  {title} {\bibinfo {title} {Grasp: A general-purpose relativistic atomic structure program},\ }\href {https://doi.org/https://doi.org/10.1016/0010-4655(89)90136-7} {\bibfield  {journal} {\bibinfo  {journal} {Comp. Phys. Comm.}\ }\textbf {\bibinfo {volume} {55}},\ \bibinfo {pages} {425} (\bibinfo {year} {1989})}\BibitemShut {NoStop}%
\bibitem [{\citenamefont {Ekman}\ \emph {et~al.}(2014)\citenamefont {Ekman}, \citenamefont {Godefroid},\ and\ \citenamefont {Hartman}}]{ekman2014validation}%
  \BibitemOpen
  \bibfield  {author} {\bibinfo {author} {\bibfnamefont {J.}~\bibnamefont {Ekman}}, \bibinfo {author} {\bibfnamefont {M.~R.}\ \bibnamefont {Godefroid}},\ and\ \bibinfo {author} {\bibfnamefont {H.}~\bibnamefont {Hartman}},\ }\bibfield  {title} {\bibinfo {title} {Validation and implementation of uncertainty estimates of calculated transition rates},\ }\href@noop {} {\bibfield  {journal} {\bibinfo  {journal} {Atoms}\ }\textbf {\bibinfo {volume} {2}},\ \bibinfo {pages} {215} (\bibinfo {year} {2014})}\BibitemShut {NoStop}%
\bibitem [{\citenamefont {Gaigalas}\ \emph {et~al.}(2020)\citenamefont {Gaigalas}, \citenamefont {Rynkun}, \citenamefont {Rad\v{z}i\={u}t\.{e}}, \citenamefont {Kato}, \citenamefont {Tanaka},\ and\ \citenamefont {J{\"o}nsson}}]{Gaigalas.2020.p13}%
  \BibitemOpen
  \bibfield  {author} {\bibinfo {author} {\bibfnamefont {G.}~\bibnamefont {Gaigalas}}, \bibinfo {author} {\bibfnamefont {P.}~\bibnamefont {Rynkun}}, \bibinfo {author} {\bibfnamefont {L.}~\bibnamefont {Rad\v{z}i\={u}t\.{e}}}, \bibinfo {author} {\bibfnamefont {D.}~\bibnamefont {Kato}}, \bibinfo {author} {\bibfnamefont {M.}~\bibnamefont {Tanaka}},\ and\ \bibinfo {author} {\bibfnamefont {P.}~\bibnamefont {J{\"o}nsson}},\ }\bibfield  {title} {\bibinfo {title} {Energy level structure and transition data of er$^{2+}$},\ }\href {https://doi.org/10.3847/1538-4365/ab881a} {\bibfield  {journal} {\bibinfo  {journal} {Astrophys. J. Suppl. Ser.}\ }\textbf {\bibinfo {volume} {248}},\ \bibinfo {pages} {13} (\bibinfo {year} {2020})}\BibitemShut {NoStop}%
\bibitem [{\citenamefont {Rynkun}\ \emph {et~al.}(2022)\citenamefont {Rynkun}, \citenamefont {Banerjee}, \citenamefont {Gaigalas}, \citenamefont {Tanaka}, \citenamefont {Radžiūtė},\ and\ \citenamefont {Kato}}]{Rynkun.2022.p82}%
  \BibitemOpen
  \bibfield  {author} {\bibinfo {author} {\bibfnamefont {P.}~\bibnamefont {Rynkun}}, \bibinfo {author} {\bibfnamefont {S.}~\bibnamefont {Banerjee}}, \bibinfo {author} {\bibfnamefont {G.}~\bibnamefont {Gaigalas}}, \bibinfo {author} {\bibfnamefont {M.}~\bibnamefont {Tanaka}}, \bibinfo {author} {\bibfnamefont {L.}~\bibnamefont {Radžiūtė}},\ and\ \bibinfo {author} {\bibfnamefont {D.}~\bibnamefont {Kato}},\ }\bibfield  {title} {\bibinfo {title} {Theoretical investigation of energy levels and transition for ce iv},\ }\href {https://doi.org/10.1051/0004-6361/202141513} {\bibfield  {journal} {\bibinfo  {journal} {A\&A}\ }\textbf {\bibinfo {volume} {658}},\ \bibinfo {pages} {A82} (\bibinfo {year} {2022})}\BibitemShut {NoStop}%
\bibitem [{\citenamefont {Gaigalas}\ \emph {et~al.}(2022)\citenamefont {Gaigalas}, \citenamefont {Rynkun}, \citenamefont {Banerjee}, \citenamefont {Tanaka}, \citenamefont {Kato},\ and\ \citenamefont {Rad\v{z}i\={u}t\.{e}}}]{Gaigalas.2022.p281}%
  \BibitemOpen
  \bibfield  {author} {\bibinfo {author} {\bibfnamefont {G.}~\bibnamefont {Gaigalas}}, \bibinfo {author} {\bibfnamefont {P.}~\bibnamefont {Rynkun}}, \bibinfo {author} {\bibfnamefont {S.}~\bibnamefont {Banerjee}}, \bibinfo {author} {\bibfnamefont {M.}~\bibnamefont {Tanaka}}, \bibinfo {author} {\bibfnamefont {D.}~\bibnamefont {Kato}},\ and\ \bibinfo {author} {\bibfnamefont {L.}~\bibnamefont {Rad\v{z}i\={u}t\.{e}}},\ }\bibfield  {title} {\bibinfo {title} {Theoretical investigation of energy levels and transitions for pr iv},\ }\href {https://doi.org/10.1093/mnras/stac2401} {\bibfield  {journal} {\bibinfo  {journal} {MNRAS}\ }\textbf {\bibinfo {volume} {517}},\ \bibinfo {pages} {281} (\bibinfo {year} {2022})}\BibitemShut {NoStop}%
\bibitem [{\citenamefont {Kitovienė}\ \emph {et~al.}(2024)\citenamefont {Kitovienė}, \citenamefont {Gaigalas}, \citenamefont {Rynkun}, \citenamefont {Tanaka},\ and\ \citenamefont {Kato}}]{Kitoviene.2024.p}%
  \BibitemOpen
  \bibfield  {author} {\bibinfo {author} {\bibfnamefont {L.}~\bibnamefont {Kitovienė}}, \bibinfo {author} {\bibfnamefont {G.}~\bibnamefont {Gaigalas}}, \bibinfo {author} {\bibfnamefont {P.}~\bibnamefont {Rynkun}}, \bibinfo {author} {\bibfnamefont {M.}~\bibnamefont {Tanaka}},\ and\ \bibinfo {author} {\bibfnamefont {D.}~\bibnamefont {Kato}},\ }\bibfield  {title} {\bibinfo {title} {Theoretical investigation of the ge isoelectronic sequence},\ }\bibfield  {journal} {\bibinfo  {journal} {J. Phys. Chem. Ref. Data}\ }\textbf {\bibinfo {volume} {53}},\ \href {https://doi.org/10.1063/5.0187307} {10.1063/5.0187307} (\bibinfo {year} {2024})\BibitemShut {NoStop}%
\bibitem [{\citenamefont {Kramida}\ \emph {et~al.}(2023)\citenamefont {Kramida}, \citenamefont {{Yu.~Ralchenko}}, \citenamefont {Reader},\ and\ \citenamefont {{and NIST ASD Team}}}]{Kramida.2023.p}%
  \BibitemOpen
  \bibfield  {author} {\bibinfo {author} {\bibfnamefont {A.}~\bibnamefont {Kramida}}, \bibinfo {author} {\bibnamefont {{Yu.~Ralchenko}}}, \bibinfo {author} {\bibfnamefont {J.}~\bibnamefont {Reader}},\ and\ \bibinfo {author} {\bibnamefont {{and NIST ASD Team}}},\ }\href@noop {} {}\bibinfo {howpublished} {{NIST Atomic Spectra Database (ver. 5.11), [Online]. Available: {\tt{https://physics.nist.gov/asd}} [2024, July 5]. National Institute of Standards and Technology, Gaithersburg, MD.}} (\bibinfo {year} {2023})\BibitemShut {NoStop}%
\bibitem [{\citenamefont {Haque}\ and\ \citenamefont {Mukherjee}(1984)}]{mukherjee84}%
  \BibitemOpen
  \bibfield  {author} {\bibinfo {author} {\bibfnamefont {M.~A.}\ \bibnamefont {Haque}}\ and\ \bibinfo {author} {\bibfnamefont {D.}~\bibnamefont {Mukherjee}},\ }\bibfield  {title} {\bibinfo {title} {Application of cluster expansion techniques to open shells: Calculation of difference energies},\ }\href {https://doi.org/10.1063/1.446574} {\bibfield  {journal} {\bibinfo  {journal} {J. Chem. Phys.}\ }\textbf {\bibinfo {volume} {80}},\ \bibinfo {pages} {5058} (\bibinfo {year} {1984})}\BibitemShut {NoStop}%
\bibitem [{\citenamefont {Lindgren}\ and\ \citenamefont {Mukherjee}(1987)}]{lindgren87}%
  \BibitemOpen
  \bibfield  {author} {\bibinfo {author} {\bibfnamefont {I.}~\bibnamefont {Lindgren}}\ and\ \bibinfo {author} {\bibfnamefont {D.}~\bibnamefont {Mukherjee}},\ }\bibfield  {title} {\bibinfo {title} {On the connectivity criteria in the open-shell coupled-cluster theory for general model spaces},\ }\href {https://doi.org/10.1016/0370-1573(87)90073-1} {\bibfield  {journal} {\bibinfo  {journal} {Phys. Rep.}\ }\textbf {\bibinfo {volume} {151}},\ \bibinfo {pages} {93–127} (\bibinfo {year} {1987})}\BibitemShut {NoStop}%
\bibitem [{\citenamefont {Lyons}\ and\ \citenamefont {Gallagher}(1998)}]{Mgionpol}%
  \BibitemOpen
  \bibfield  {author} {\bibinfo {author} {\bibfnamefont {B.~J.}\ \bibnamefont {Lyons}}\ and\ \bibinfo {author} {\bibfnamefont {T.~F.}\ \bibnamefont {Gallagher}},\ }\bibfield  {title} {\bibinfo {title} {Mg $3snf\ensuremath{-}3sng\ensuremath{-}3snh\ensuremath{-}3sni$ intervals and the ${\mathrm{mg}}^{+}$ dipole polarizability},\ }\href {https://doi.org/10.1103/PhysRevA.57.2426} {\bibfield  {journal} {\bibinfo  {journal} {Phys. Rev. A}\ }\textbf {\bibinfo {volume} {57}},\ \bibinfo {pages} {2426} (\bibinfo {year} {1998})}\BibitemShut {NoStop}%
\bibitem [{\citenamefont {Xu}\ \emph {et~al.}(2017)\citenamefont {Xu}, \citenamefont {Deng}, \citenamefont {Che}, \citenamefont {Yuan}, \citenamefont {Zhang},\ and\ \citenamefont {Lu}}]{Xu2017}%
  \BibitemOpen
  \bibfield  {author} {\bibinfo {author} {\bibfnamefont {Z.~T.}\ \bibnamefont {Xu}}, \bibinfo {author} {\bibfnamefont {K.}~\bibnamefont {Deng}}, \bibinfo {author} {\bibfnamefont {H.}~\bibnamefont {Che}}, \bibinfo {author} {\bibfnamefont {W.~H.}\ \bibnamefont {Yuan}}, \bibinfo {author} {\bibfnamefont {J.}~\bibnamefont {Zhang}},\ and\ \bibinfo {author} {\bibfnamefont {Z.~H.}\ \bibnamefont {Lu}},\ }\bibfield  {title} {\bibinfo {title} {Precision measurement of the $^{25}\mathrm{Mg}^{+}$ ground-state hyperfine constant},\ }\href {https://doi.org/10.1103/PhysRevA.96.052507} {\bibfield  {journal} {\bibinfo  {journal} {Phys. Rev. A}\ }\textbf {\bibinfo {volume} {96}},\ \bibinfo {pages} {052507} (\bibinfo {year} {2017})}\BibitemShut {NoStop}%
\bibitem [{\citenamefont {Itano}\ and\ \citenamefont {Wineland}(1981)}]{Itano1981}%
  \BibitemOpen
  \bibfield  {author} {\bibinfo {author} {\bibfnamefont {W.~M.}\ \bibnamefont {Itano}}\ and\ \bibinfo {author} {\bibfnamefont {D.~J.}\ \bibnamefont {Wineland}},\ }\bibfield  {title} {\bibinfo {title} {Precision measurement of the ground-state hyperfine constant of $^{25}\mathrm{Mg}^{+}$},\ }\href {https://doi.org/10.1103/PhysRevA.24.1364} {\bibfield  {journal} {\bibinfo  {journal} {Phys. Rev. A}\ }\textbf {\bibinfo {volume} {24}},\ \bibinfo {pages} {1364} (\bibinfo {year} {1981})}\BibitemShut {NoStop}%
\bibitem [{\citenamefont {Nandy}\ and\ \citenamefont {Sahoo}(2014)}]{dillip2014}%
  \BibitemOpen
  \bibfield  {author} {\bibinfo {author} {\bibfnamefont {D.~K.}\ \bibnamefont {Nandy}}\ and\ \bibinfo {author} {\bibfnamefont {B.~K.}\ \bibnamefont {Sahoo}},\ }\bibfield  {title} {\bibinfo {title} {Quadrupole shifts for the $^{171}\mathrm{Yb}^{+}$ ion clocks: Experiments versus theories},\ }\href {https://doi.org/10.1103/PhysRevA.90.050503} {\bibfield  {journal} {\bibinfo  {journal} {Phys. Rev. A}\ }\textbf {\bibinfo {volume} {90}},\ \bibinfo {pages} {050503} (\bibinfo {year} {2014})}\BibitemShut {NoStop}%
\bibitem [{\citenamefont {Sahoo}\ and\ \citenamefont {Ohayon}(2021)}]{bijaya-li}%
  \BibitemOpen
  \bibfield  {author} {\bibinfo {author} {\bibfnamefont {B.~K.}\ \bibnamefont {Sahoo}}\ and\ \bibinfo {author} {\bibfnamefont {B.}~\bibnamefont {Ohayon}},\ }\bibfield  {title} {\bibinfo {title} {Benchmarking many-body approaches for the determination of isotope-shift constants: Application to the $\mathrm{Li}, {\mathrm{be}}^{+}$, and ${\mathrm{ar}}^{15+}$ isoelectronic systems},\ }\href {https://doi.org/10.1103/PhysRevA.103.052802} {\bibfield  {journal} {\bibinfo  {journal} {Phys. Rev. A}\ }\textbf {\bibinfo {volume} {103}},\ \bibinfo {pages} {052802} (\bibinfo {year} {2021})}\BibitemShut {NoStop}%
\bibitem [{\citenamefont {Papoulia}\ \emph {et~al.}(2019)\citenamefont {Papoulia}, \citenamefont {Ekman}, \citenamefont {Gaigalas}, \citenamefont {Godefroid}, \citenamefont {Gustafsson}, \citenamefont {Hartman}, \citenamefont {Li}, \citenamefont {Radžiūtė}, \citenamefont {Rynkun}, \citenamefont {Schiffmann}, \citenamefont {Wang},\ and\ \citenamefont {J\"onsson}}]{atoms7040106}%
  \BibitemOpen
  \bibfield  {author} {\bibinfo {author} {\bibfnamefont {A.}~\bibnamefont {Papoulia}}, \bibinfo {author} {\bibfnamefont {J.}~\bibnamefont {Ekman}}, \bibinfo {author} {\bibfnamefont {G.}~\bibnamefont {Gaigalas}}, \bibinfo {author} {\bibfnamefont {M.}~\bibnamefont {Godefroid}}, \bibinfo {author} {\bibfnamefont {S.}~\bibnamefont {Gustafsson}}, \bibinfo {author} {\bibfnamefont {H.}~\bibnamefont {Hartman}}, \bibinfo {author} {\bibfnamefont {W.}~\bibnamefont {Li}}, \bibinfo {author} {\bibfnamefont {L.}~\bibnamefont {Radžiūtė}}, \bibinfo {author} {\bibfnamefont {P.}~\bibnamefont {Rynkun}}, \bibinfo {author} {\bibfnamefont {S.}~\bibnamefont {Schiffmann}}, \bibinfo {author} {\bibfnamefont {K.}~\bibnamefont {Wang}},\ and\ \bibinfo {author} {\bibfnamefont {P.}~\bibnamefont {J\"onsson}},\ }\bibfield  {title} {\bibinfo {title} {Coulomb (velocity) gauge recommended in multiconfiguration calculations of transition data involving rydberg series},\ }\bibfield  {journal} {\bibinfo  {journal} {Atoms}\ }\textbf {\bibinfo
  {volume} {7}},\ \href {https://doi.org/10.3390/atoms7040106} {10.3390/atoms7040106} (\bibinfo {year} {2019})\BibitemShut {NoStop}%
\bibitem [{\citenamefont {Pehlivan~Rhodin}\ \emph {et~al.}(2017)\citenamefont {Pehlivan~Rhodin}, \citenamefont {Hartman}, \citenamefont {Nilsson},\ and\ \citenamefont {J\"onsson}}]{Pehlivan}%
  \BibitemOpen
  \bibfield  {author} {\bibinfo {author} {\bibfnamefont {A.}~\bibnamefont {Pehlivan~Rhodin}}, \bibinfo {author} {\bibfnamefont {H.}~\bibnamefont {Hartman}}, \bibinfo {author} {\bibfnamefont {H.}~\bibnamefont {Nilsson}},\ and\ \bibinfo {author} {\bibfnamefont {P.}~\bibnamefont {J\"onsson}},\ }\bibfield  {title} {\bibinfo {title} {Experimental and theoretical oscillator strengths of mg i for accurate abundance analysis},\ }\bibfield  {journal} {\bibinfo  {journal} {A\&A}\ }\textbf {\bibinfo {volume} {598}},\ \href {https://doi.org/10.1051/0004-6361/201629849} {10.1051/0004-6361/201629849} (\bibinfo {year} {2017})\BibitemShut {NoStop}%
\bibitem [{\citenamefont {Schiffmann}\ \emph {et~al.}(2020)\citenamefont {Schiffmann}, \citenamefont {Godefroid}, \citenamefont {Ekman}, \citenamefont {J\"onsson},\ and\ \citenamefont {Fischer}}]{PhysRevA.101.062510}%
  \BibitemOpen
  \bibfield  {author} {\bibinfo {author} {\bibfnamefont {S.}~\bibnamefont {Schiffmann}}, \bibinfo {author} {\bibfnamefont {M.}~\bibnamefont {Godefroid}}, \bibinfo {author} {\bibfnamefont {J.}~\bibnamefont {Ekman}}, \bibinfo {author} {\bibfnamefont {P.}~\bibnamefont {J\"onsson}},\ and\ \bibinfo {author} {\bibfnamefont {C.~F.}\ \bibnamefont {Fischer}},\ }\bibfield  {title} {\bibinfo {title} {Natural orbitals in multiconfiguration calculations of hyperfine-structure parameters},\ }\href {https://doi.org/10.1103/PhysRevA.101.062510} {\bibfield  {journal} {\bibinfo  {journal} {Phys. Rev. A}\ }\textbf {\bibinfo {volume} {101}},\ \bibinfo {pages} {062510} (\bibinfo {year} {2020})}\BibitemShut {NoStop}%
\bibitem [{\citenamefont {Si}\ \emph {et~al.}(2025)\citenamefont {Si}, \citenamefont {Li}, \citenamefont {Wang}, \citenamefont {Chen}, \citenamefont {Gaigalas}, \citenamefont {Godefroid},\ and\ \citenamefont {Jönsson}}]{SI2025109604}%
  \BibitemOpen
  \bibfield  {author} {\bibinfo {author} {\bibfnamefont {R.}~\bibnamefont {Si}}, \bibinfo {author} {\bibfnamefont {Y.}~\bibnamefont {Li}}, \bibinfo {author} {\bibfnamefont {K.}~\bibnamefont {Wang}}, \bibinfo {author} {\bibfnamefont {C.}~\bibnamefont {Chen}}, \bibinfo {author} {\bibfnamefont {G.}~\bibnamefont {Gaigalas}}, \bibinfo {author} {\bibfnamefont {M.}~\bibnamefont {Godefroid}},\ and\ \bibinfo {author} {\bibfnamefont {P.}~\bibnamefont {Jönsson}},\ }\bibfield  {title} {\bibinfo {title} {Graspg – an extension to grasp2018 based on configuration state function generators},\ }\href {https://doi.org/https://doi.org/10.1016/j.cpc.2025.109604} {\bibfield  {journal} {\bibinfo  {journal} {Comp. Phys. Comm.}\ }\textbf {\bibinfo {volume} {312}},\ \bibinfo {pages} {109604} (\bibinfo {year} {2025})}\BibitemShut {NoStop}%
\bibitem [{\citenamefont {Froese~Fischer}\ \emph {et~al.}(2019)\citenamefont {Froese~Fischer}, \citenamefont {Gaigalas}, \citenamefont {Jönsson},\ and\ \citenamefont {Bieroń}}]{FroeseFischer.2019.p184}%
  \BibitemOpen
  \bibfield  {author} {\bibinfo {author} {\bibfnamefont {C.}~\bibnamefont {Froese~Fischer}}, \bibinfo {author} {\bibfnamefont {G.}~\bibnamefont {Gaigalas}}, \bibinfo {author} {\bibfnamefont {P.}~\bibnamefont {Jönsson}},\ and\ \bibinfo {author} {\bibfnamefont {J.}~\bibnamefont {Bieroń}},\ }\bibfield  {title} {\bibinfo {title} {Grasp2018—a fortran 95 version of the general relativistic atomic structure package},\ }\href {https://doi.org/10.1016/j.cpc.2018.10.032} {\bibfield  {journal} {\bibinfo  {journal} {Comp. Phys. Comm.}\ }\textbf {\bibinfo {volume} {237}},\ \bibinfo {pages} {184} (\bibinfo {year} {2019})}\BibitemShut {NoStop}%
\bibitem [{\citenamefont {Schiffmann}\ \emph {et~al.}(2022)\citenamefont {Schiffmann}, \citenamefont {Li}, \citenamefont {Ekman}, \citenamefont {Gaigalas}, \citenamefont {Godefroid}, \citenamefont {Jönsson},\ and\ \citenamefont {Bieroń}}]{SCHIFFMANN2022108403}%
  \BibitemOpen
  \bibfield  {author} {\bibinfo {author} {\bibfnamefont {S.}~\bibnamefont {Schiffmann}}, \bibinfo {author} {\bibfnamefont {J.}~\bibnamefont {Li}}, \bibinfo {author} {\bibfnamefont {J.}~\bibnamefont {Ekman}}, \bibinfo {author} {\bibfnamefont {G.}~\bibnamefont {Gaigalas}}, \bibinfo {author} {\bibfnamefont {M.}~\bibnamefont {Godefroid}}, \bibinfo {author} {\bibfnamefont {P.}~\bibnamefont {Jönsson}},\ and\ \bibinfo {author} {\bibfnamefont {J.}~\bibnamefont {Bieroń}},\ }\bibfield  {title} {\bibinfo {title} {Relativistic radial electron density functions and natural orbitals from grasp2018},\ }\href {https://doi.org/https://doi.org/10.1016/j.cpc.2022.108403} {\bibfield  {journal} {\bibinfo  {journal} {Comp. Phys. Comm.}\ }\textbf {\bibinfo {volume} {278}},\ \bibinfo {pages} {108403} (\bibinfo {year} {2022})}\BibitemShut {NoStop}%
\bibitem [{\citenamefont {Fischer}\ \emph {et~al.}(1997)\citenamefont {Fischer}, \citenamefont {Brage},\ and\ \citenamefont {J{\"o}nsson}}]{MCHF}%
  \BibitemOpen
  \bibfield  {author} {\bibinfo {author} {\bibfnamefont {C.~F.}\ \bibnamefont {Fischer}}, \bibinfo {author} {\bibfnamefont {T.}~\bibnamefont {Brage}},\ and\ \bibinfo {author} {\bibfnamefont {P.}~\bibnamefont {J{\"o}nsson}},\ }\bibfield  {title} {\bibinfo {title} {Computational atomic structure: An mchf approach}\ }(\bibinfo {year} {1997})\BibitemShut {NoStop}%
\bibitem [{\citenamefont {Yordanov}\ \emph {et~al.}(2012)\citenamefont {Yordanov}, \citenamefont {Bissell}, \citenamefont {Blaum}, \citenamefont {De~Rydt}, \citenamefont {Geppert}, \citenamefont {Kowalska}, \citenamefont {Kr\"amer}, \citenamefont {Kreim}, \citenamefont {Krieger}, \citenamefont {Lievens}, \citenamefont {Neff}, \citenamefont {Neugart}, \citenamefont {Neyens}, \citenamefont {N\"ortersh\"auser}, \citenamefont {S\'anchez},\ and\ \citenamefont {Vingerhoets}}]{2012-Yor}%
  \BibitemOpen
  \bibfield  {author} {\bibinfo {author} {\bibfnamefont {D.~T.}\ \bibnamefont {Yordanov}}, \bibinfo {author} {\bibfnamefont {M.~L.}\ \bibnamefont {Bissell}}, \bibinfo {author} {\bibfnamefont {K.}~\bibnamefont {Blaum}}, \bibinfo {author} {\bibfnamefont {M.}~\bibnamefont {De~Rydt}}, \bibinfo {author} {\bibfnamefont {C.}~\bibnamefont {Geppert}}, \bibinfo {author} {\bibfnamefont {M.}~\bibnamefont {Kowalska}}, \bibinfo {author} {\bibfnamefont {J.}~\bibnamefont {Kr\"amer}}, \bibinfo {author} {\bibfnamefont {K.}~\bibnamefont {Kreim}}, \bibinfo {author} {\bibfnamefont {A.}~\bibnamefont {Krieger}}, \bibinfo {author} {\bibfnamefont {P.}~\bibnamefont {Lievens}}, \bibinfo {author} {\bibfnamefont {T.}~\bibnamefont {Neff}}, \bibinfo {author} {\bibfnamefont {R.}~\bibnamefont {Neugart}}, \bibinfo {author} {\bibfnamefont {G.}~\bibnamefont {Neyens}}, \bibinfo {author} {\bibfnamefont {W.}~\bibnamefont {N\"ortersh\"auser}}, \bibinfo {author} {\bibfnamefont {R.}~\bibnamefont {S\'anchez}},\ and\ \bibinfo {author} {\bibfnamefont
  {P.}~\bibnamefont {Vingerhoets}},\ }\bibfield  {title} {\bibinfo {title} {{Nuclear Charge Radii of $^{21\mathrm{\text{\ensuremath{-}}}32}\mathrm{Mg}$}},\ }\href {https://doi.org/10.1103/PhysRevLett.108.042504} {\bibfield  {journal} {\bibinfo  {journal} {Phys. Rev. Lett.}\ }\textbf {\bibinfo {volume} {108}},\ \bibinfo {pages} {042504} (\bibinfo {year} {2012})}\BibitemShut {NoStop}%
\bibitem [{\citenamefont {Batteiger}\ \emph {et~al.}(2009)\citenamefont {Batteiger}, \citenamefont {Kn\"unz}, \citenamefont {Herrmann}, \citenamefont {Saathoff}, \citenamefont {Sch\"ussler}, \citenamefont {Bernhardt}, \citenamefont {Wilken}, \citenamefont {Holzwarth}, \citenamefont {H\"ansch},\ and\ \citenamefont {Udem}}]{2009-MgTrap}%
  \BibitemOpen
  \bibfield  {author} {\bibinfo {author} {\bibfnamefont {V.}~\bibnamefont {Batteiger}}, \bibinfo {author} {\bibfnamefont {S.}~\bibnamefont {Kn\"unz}}, \bibinfo {author} {\bibfnamefont {M.}~\bibnamefont {Herrmann}}, \bibinfo {author} {\bibfnamefont {G.}~\bibnamefont {Saathoff}}, \bibinfo {author} {\bibfnamefont {H.~A.}\ \bibnamefont {Sch\"ussler}}, \bibinfo {author} {\bibfnamefont {B.}~\bibnamefont {Bernhardt}}, \bibinfo {author} {\bibfnamefont {T.}~\bibnamefont {Wilken}}, \bibinfo {author} {\bibfnamefont {R.}~\bibnamefont {Holzwarth}}, \bibinfo {author} {\bibfnamefont {T.~W.}\ \bibnamefont {H\"ansch}},\ and\ \bibinfo {author} {\bibfnamefont {T.}~\bibnamefont {Udem}},\ }\bibfield  {title} {\bibinfo {title} {Precision spectroscopy of the $3s\text{\ensuremath{-}}3p$ fine-structure doublet in {Mg}$^{+}$},\ }\href {https://doi.org/10.1103/PhysRevA.80.022503} {\bibfield  {journal} {\bibinfo  {journal} {Phys. Rev. A}\ }\textbf {\bibinfo {volume} {80}},\ \bibinfo {pages} {022503} (\bibinfo {year} {2009})}\BibitemShut
  {NoStop}%
\end{thebibliography}%

\end{document}